\documentclass[epj]{svjour}
\usepackage{graphicx,wrapfig,lipsum}
\usepackage{amssymb,amsmath}\usepackage{color}
\usepackage{subeqnarray}
\usepackage[utf8]{inputenc}
\usepackage{url,hyperref,doi} 
\makeatletter
\g@addto@macro{\UrlBreaks}{\UrlOrds}
\makeatother
 
\begin{document}
\title{Magnetic Dipole Moment in Relativistic Quantum Mechanics}
\author{Andrew Steinmetz, Martin Formanek, and Johann Rafelski}
%
%
\institute{Department of Physics, The University of Arizona, 
Tucson, Arizona, 85721, USA}
\date{Submitted:November 30, 2018 / Print date: \today}
%
\abstract{We investigate relativistic quantum mechanics (RQM) for particles with arbitrary magnetic moment. We compare two well known RQM models: a) Dirac equation supplemented with an incremental Pauli term (DP); b) Klein-Gordon equations with full Pauli EM dipole moment term (KGP). We compare exact solutions to the external field cases in the limit of weak and strong (critical) fields for: i) homogeneous magnetic field, and ii) the Coulomb $1/r$-potential. For i) we consider the Landau energies and the Landau states as a function of the gyromagnetic factor ($g$-factor). For ii) we investigate contribution to the Lamb shift and the fine structure splitting. For both we address the limit of strong binding and show that these two formulations grossly disagree. We discuss possible experiments capable of distinguishing between KGP and DP models in laboratory. We describe impact of our considerations in astrophysical context (magnetars). We introduce novel RQM models of magnetic moments which can be further explored.
\PACS{%
 {13.40.Em}{Electric and magnetic moments}
 {03.65.Pm}{Relativistic wave equations}
 {31.30.J-}{Relativistic and quantum electrodynamics (QED)}
 {36.10.-k}{Exotic atoms and molecules}
 {97.60.Gb}{Pulsars, Magnetars}
 } 
} 
\authorrunning{A. Steinmetz, M. Formanek and J. Rafelski}
\titlerunning{Magnetic Dipole Moment in Relativistic QM}
\maketitle

\section{Introduction} \label{pre_intro}

\subsection{Magnetic Moment in QM} \label{intro}
We have recently recognized~\cite{Rafelski:2017hce} that an additional physical principle or guiding assumption is required to uniquely characterize relativistic {\em classical} particle dynamics with magnetic moment. In the current work we address the {\em quantum dynamics} non-uniqueness~\cite{Brown:1958zz,Veltman:1997am}.

In the following we study relativistic Fermi particles with \lq anomalous\rq\ $a\ne 0$ gyromagnetic ratio $g$ {\it i.e.\/} anomalous magnetic moment (AMM) 
\begin{alignat}{1}
\label{intro01} &\vec{\mu}={g}\,\frac{e\hbar}{2mc}\,\frac{\vec{\sigma}}{2}
\equiv-g\,\mu_{B}\,\frac{\vec{\sigma}}{2}\;,
\quad a\equiv \frac g 2-1\;,
\end{alignat}
where $\mu_{B}$ is the Bohr magneton\footnote[1]{We will use the conventions of negative electric charge $e=-|e|$ and Gaussian units throughout.}.

We address two different models of introducing anomalous magnetic moment in QM: 
\begin{enumerate}
\item[(a)] the Dirac-Pauli (DP) first order equation which is the Dirac equation where $g$-factor is precisely fixed to the $g=2$, with the addition of an incremental Pauli term; and
\item[(a)] the Klein-Gordon-Pauli (KGP) second order equation which \lq\lq squares\rq\rq\ the Dirac equation and thereafter allows the magnetic moment $\vec{\mu}$ to vary independently of charge and mass, unlike Dirac theory.
\end{enumerate} 
These two approaches coincide when the anomaly $a$ vanishes. However, all particles that have magnetic moments differ from the Dirac value $g=2$, either due to their composite nature, or, for point particles, due to the quantum vacuum fluctuation effect.

We find that even a small magnetic anomaly has a large effect in the limit of strong fields generated by massive magnetar stars~\cite{Kaspi:2017fwg}. Therefore it is not clear that the tacit assumption of $g=2$ in the case of strong fields~\cite{Rafelski:1976ts,Greiner:1985ce,Rafelski:2016ixr} is allowed~\cite{Evans:2018kor}. This argument is especially strong considering tightly bound composite particles such as protons and neutrons where the large anomalous magnetic moment can be taken as an external prescribed parameter unrelated to the elementary quantum vacuum fluctuations, and it is of particular interest to study the dynamical behavior of these particles in fields of magnetar strength. This interest carries over to the environment of strong fields created in focus of ultra-intense laser pulses and the associated particle production processes~\cite{Dunne:2014qda,Hegelich:2014tda}. We consider also precision spectroscopic experiments and recognize consequences even in the weak coupling limit. We evaluate contributions for precision hydrogen muonic atom Lamb shift experiments~\cite{Pohl:2013yb}. 

All the above provides motivation for a closer analysis of how different methods of introducing anomalous magnetic moment to RQM lead to different physical predictions, which allows model elimination by comparison with experiment. While we will discuss consequences of the previously explored DP equation, the majority of the novel material in this paper will deal with the KGP equation which has received less attention in prior literature and thus, for purpose of this work, most of the results need to be newly derived, except where noted otherwise.

One of the important outcomes of this work is the recognition of additional selection criteria for effective field theory (EFT) approaches offered by the presence of strong fields. In the weak field environment, both methods, KGP and DP, are suitable as EFT methods of dealing with magnetic moment. Our results show vital differences in the physical outcomes in presence of strong fields. We will argue that the DP version is not favored and that the KGP-EFT approach may need a further improvement. In our work we address solely point particles, thus challenges of Zemach moments \cite{Zemach:1956zz} do not enter the present discussion. 

The advantage of the the KGP dynamics is that it incorporates both possible signs of the magnetic moment, irrespective of the sign of the electric charge, a choice that is lost in the DP single particle model. The magnetic moment sign choice is tacitly made when writing for the Dirac operator $\gamma\cdot p \pm m$; choosing minus sign in the mass term we choose the conventional magnetic moment sign, the other sign leads to opposite magnetic moment sign. The incorporation of both magnetic moment signs in fundamental equation for relativistic fermion dynamics doubles the number of degrees of freedom from usual 4 in the Dirac case to $2\times 4=8$ in the KGP case.

Some may ask why the behavior of effective single particle model theories in the presence of strong external fields warrants our interest. A foundational motivation to offer the present study is that we believe it has never been settled which quantum theory description should be the basis for the development of the quantum field theory of matter and fields: for example Schwinger in the study of Euler-Heisenberg-Schwinger effective EM action~\cite{Schwinger:1951nm} computes the vacuum fluctuations adopting as basis the KGP dynamics, followed by factor 1/2 to remove the magnetic moment sign degeneracy as both particles with positive and negative magnetic moment were allowed to fluctuate in the vacuum.

\subsection{Outline} \label{outline}
We introduce in quantitative fashion relativistic quantum equations with anomalous magnetic moment in section~\ref{RQMg}, presenting both DP (section~\ref{dp}) and KGP (section~\ref{kgp}) equations. Relativistic covariance of both KGP and DP equations is addressed in section \ref{sym}. In section~\ref{currents} We discuss currents and the physical degrees of freedom inherent to the higher order formulation in section~\ref{freedom}. This also allows us to comment on the procedure of canonical 2nd quantization, but the full treatment is beyond the scope of the present work.

We then turn in section~\ref{lan} to study the case of a homogeneous magnetic field. We solve the KGP equation in section~\ref{lankpg}) and analyze the Landau levels using the ladder operator method in section~\ref{ladder}. We show in section~\ref{lannonrel} that, like the nonrelativistic case, the Landau levels lose their degeneracy in the presence of an anomalous magnetic moment, but that this degeneracy is restored for certain values of the $g$-factor. We compare these results to the DP case in section~\ref{DPBhom}, relying on solutions available in literature~\cite{Tsai:1972iq}. We argue that DP solutions, which do not depend only on the magnetic moment of a particle but in an incoherent way also on anomaly, lack the required physical simplicity we see in the KGP case.

In section~\ref{cou} we solve the KGP-Coulomb problem and report the exact energy levels along with the perturbative expansion showing the influence of the (anomalous) magnetic moment on the fine structure and Lamb shift which differ from the expressions obtained in Dirac ($g=2$) or DP formulations. We evaluate the hydrogen and certain exotic atom effects, searching for an opportunity for further experimental insight in modern spectroscopic experiments.

Section~\ref{sb} deals with the consequences of strong binding exhibited in critical high strength fields. The appropriateness of using the DP and KGP equations as effective theories is discussed in section~\ref{vacfl}. The case of extreme magnetic fields, such as those found in magnetars, is considered in section~\ref{sbl}. Section~\ref{IKGP} suggests an improved version of the KGP equation, which produces better strong field behavior. For critical binding of high-$Z$ nuclei we compare in section~\ref{sbc} the analytic solutions of the KGP equation to the numerical solutions of DP presented by Thaller~\cite{Thaller:1992ji}, and review another analysis done by Barut and Kraus~\cite{Barut:1975hz,Barut:1976hs}. Our findings are summarized in section~\ref{concl} where we also discuss future research directions.

\section{Relativistic equation with anomalous magnetic moment}\label{RQMg}
\subsection{Dirac-Pauli equation} \label{dp}
The DP formulation is given by the Dirac equation with an additional Pauli term that is responsible for the anomalous moment
\begin{alignat}{1}
\label{dp01} &\left(\gamma^{\mu}\left(i\hbar c\partial_{\mu}-eA_{\mu}\right)-mc^{2}-a\frac{e\hbar}{4mc}\sigma_{\mu\nu}F^{\mu\nu}\right)\psi=0\;,
\end{alignat}
where\\[-1.2cm]
\begin{alignat}{1}
\label{dp02} &\sigma_{\mu\nu}=\frac{i}{2}[\gamma_{\mu},\gamma_{\nu}]
\;,\end{alignat}
is the spin tensor that is proportional to the commutator of the gamma matrices and $F^{\mu\nu}$ is the standard EM field tensor. Unless otherwise stated, all gamma matrices will be evaluated in the Dirac representation. 

The last term in Eq.\,\eqref{dp01}, known as the Pauli term, represents a coupling of magnetic moment to electric and magnetic fields and evaluates to
\begin{alignat}{1}
\label{dp03} &\frac{1}{2} \sigma_{\mu\nu}F^{\mu\nu} =i\vec{\alpha}\cdot\vec{E}-\vec{\Sigma}\cdot\vec{B}\;,
\end{alignat}
where we use the conventions
\begin{alignat}{1}
\label{dp04} & 
 \vec{\alpha}=\gamma^{0}\vec{\gamma}\;,\ 
 \vec{\Sigma}=\gamma^{5} \vec{\alpha}\;, \ 
 \gamma_5=i\gamma_0\gamma_1\gamma_2\gamma_3\;, \ 
 \gamma_5^2=1 \;.
\end{alignat} 
Equation\,\eqref{dp01} in the non-relativistic limit incorporates the Hamiltonian term
\begin{alignat}{1}
\label{dp06} &\hat{H}_{\mathrm{AMM}}=-a\frac{e\hbar}{2mc}\vec{\sigma}\cdot\vec{B}\;,\end{alignat}
which is the energy due to an anomalous magnetic moment. One refers here to the Shr\"{o}dinger-Pauli (SP) equation.

The popularity of Eq.\,\eqref{dp01} is due to its close connection to the Dirac equation. Moreover, DP allows for small anomalous moments perturbative exploration of any problem which has known Dirac equation solutions. The disadvantages are also known: 
\begin{enumerate}
\item 
The total magnetic moment is obscured having been split into an explicit anomalous contribution; and the implicit contribution buried in the spinor structure of the DP equation. This is not surprising given the \emph{ad-hoc} nature of introducing anomalous magnetic moment in this way. 
\item 
This formulation is barred from being the basis for a well behaved quantum field perturbative treatment; the interaction Lagrangian, which produces the anomalous moment,
\begin{alignat}{1}
\label{dp07} &\mathcal{L}_{AMM}=-a\frac{e\hbar}{4mc}\bar{\psi}\sigma_{\mu\nu}\psi F^{\mu\nu}\;,\end{alignat}
is non renormalizable~\cite{Knecht:2003kc} as the coupling coefficient has natural units of length.
\end{enumerate}

\subsection{Klein-Gordon-Pauli equation} \label{kgp}
The KGP equation, unlike the Dirac or its DP extension, is second order in derivatives 
\begin{alignat}{1}
\label{kgp01} &\left(\left(i\hbar c\partial_{\mu}-eA_{\mu}\right)^{2}-m^{2}c^{4}-\frac{g}{2}\frac{e\hbar c}{2}\sigma^{\mu\nu}F_{\mu\nu}\right)\Psi=0\;.
\end{alignat} 
For $g=2$ one finds Eq.\,\eqref{kgp01} by first considering the Dirac equation
\begin{alignat}{1}
\label{kgp02} &\mathfrak{D}_{+}\psi=\left(\gamma^{\mu}\left(i\hbar c\partial_{\mu}-eA_{\mu}\right)-mc^{2}\right)\psi=0\;.
\end{alignat}
The operator $\mathfrak{D}_{+}$ is considered to be the \lq\lq positive mass\rq\rq\ operator which vanishes on a wave function with positive mass eigenstate. We can then \lq\lq square\rq\rq\ Eq.\,\eqref{kgp02} by introducing the substitution
\begin{alignat}{1}
\label{kgp03} \psi&\rightarrow\mathfrak{D}_{-}\Psi,\\ 
\label{kgp03b} \mathfrak{D}_{-}&=\left(\gamma^{\mu}\left(i\hbar c\partial_{\mu}-eA_{\mu}\right)+mc^{2}\right)\;,\end{alignat}
into Eq.\,\eqref{kgp02}, yielding
\begin{alignat}{1}
\label{kgp04} &\mathfrak{D}_{+}\mathfrak{D}_{-}\Psi=0\;,
\end{alignat}
where $\mathfrak{D}_{-}$ is the corresponding negative mass operator. This operator is related to the traditional positive mass operator via chiral operator $\gamma^{5}$ conjugation
\begin{alignat}{1}
\label{kgp05} &\mathfrak{D}_{-}=-\gamma^{5}\mathfrak{D}_{+}\gamma^{5},\ [\mathfrak{D}_{+},\mathfrak{D}_{-}]=0\;.
\end{alignat}

We should point out that the substitution Eq.\,\eqref{kgp03} requires some coefficient of proportionality to preserve the required units, which will be discussed further in section~\ref{currents} when normalizing states of $\Psi$. Making use of the commutation and anti-commutation relations of the gamma matrices, it is easy to verify that Eq.\,\eqref{kgp04} is the KGP Eq.\,\eqref{kgp01} for $g=2$. 

The KGP formulation was introduced by Fock~\cite{Fock:1937dy} (however, with $g=2$) and subsequently studied by Feynman and Gell-Mann~\cite{Feynman:1958ty}, Brown~\cite{Brown:1958zz}, in the context of weak interactions. This interest is motivated by the observation that all operators in the wave equation Eq.\,\eqref{kgp01} commute with $\gamma^{5}$ therefore all eigenfunctions are good chiral eigenstates. 

On first sight parity seems to be violated by the Pauli term, since it does not commute with $\gamma^{0}$. However, the parity of EM fields is opposite, and this behavior assures that actual solutions are also parity eigenfunctions. This is seen more explicitly by writing Eq.\,\eqref{dp03} in the format
\begin{alignat}{1}
\label{kgp06} &\frac{1}{2}\sigma^{\mu\nu}F_{\mu\nu}=\vec{\sigma}\cdot\left(i\gamma^{5}\vec{E}-\vec{B}\right)\;.
\end{alignat}

The presence in Eq.\,\eqref{kgp01} of a single (Pauli) term which contains the entire spin and thus magnetic moment behavior of the particle makes the separation of charge and magnetic dipole moment more natural. By allowing the $g$-factor to vary to any value we are able to introduce the anomalous magnetic moment in the context of a second order equation.

The price we pay in the KGP case is losing the connection to the first order Dirac equation and the simple conjugation properties of Eq.\,\eqref{kgp05}. In other words, it is no longer obvious how to convert back to a first order equation without considering some other line of thought. Equation\,\eqref{kgp01} takes on the form of an inhomogeneous Klein-Gordon equation acting on a four-spinor field.

We can show that Eq.\,\eqref{kgp01} is mathematically distinct from Eq.\,\eqref{dp01} by attempting to \lq\lq square\rq\rq\ Eq.\,\eqref{dp01}, which produces cross terms between momentum and the Pauli term which are absent in Eq.\,\eqref{kgp01}. Efforts have also been made exploring KGP as the basis of a quantum field theory and its renormalization~\cite{Cortes:1992wr,AngelesMartinez:2011nt}. KGP has also found use as an effective field theory providing a description of the Compton scattering of low energy protons~\cite{DelgadoAcosta:2010nx}.

It is of relevance to always remember that the KGP equation remains a 4-spinor equation. We can introduce two 2-spinors which will be of immediate use when we normalize the KGP solutions we will study. Specifically, the stationary wave function $\Psi_{E}(\vec x)$ can be cast into the format
\begin{equation}
\Psi_{E}=
\left(\begin{matrix}
 \chi_E^s\\[0.12cm]
 \phi_E^s
\end{matrix}\right)\;.
\label{Spinor1KGP}
\end{equation}
Here both $\chi_E^s$ and $\phi_E^s$ are two-component spinors. For completeness let us restate the stationary form of KGP equation in this notation
 \begin{equation}
\left(\begin{matrix}
K_\mathrm{KG}+ \vec s_g\cdot \vec B & -i\vec s_g\cdot \vec E\\
 -i\vec s_g\cdot \vec E & K_\mathrm{KG}+ \vec s_g\cdot \vec B 
\end{matrix}\right)
\left(\begin{matrix}
 \chi_E^s\\[0.12cm]
 \phi_E^s
\end{matrix}\right)
=0\;,
\label{Spinor2KGP}
\end{equation}
where we introduced
 \begin{alignat}{1}
K_\mathrm{KG}&\equiv (E-eV)^2-(i\hbar c\vec \nabla - e\vec A)^2-m^2c^4\;, \\
\vec s_g& \equiv ge c \frac{\hbar\vec \sigma}{2}\;.
\end{alignat}
This form Eq.\,\eqref{Spinor2KGP} shows explicitly that the Coulomb problem involves the full 4-spinor dynamics, while the magnetic field problem naturally separates into upper and lower components. 

It is possible to prediagonalize and separate the 4-spinor format into two 2-spinor equations, which is Feynman\rq s preferred form~\cite{Feynman:1958ty}. First we break $\Psi$ into upper and lower components as seen in Eq.\,\eqref{Spinor2KGP} allowing the rewriting of the KGP equation as a set of two coupled equations for $\chi^s_E, \phi^s_E $. Then we add and subtract the two equations yielding
\begin{alignat}{1}
\label{lan32} 0&\!=\!\left(\left(i\hbar c\partial_{\mu}-eA_{\mu}\right)^{2}\!\!-m^{2}c^{4}\!\! +\vec s_g\cdot\vec{F}_{\pm}\right)\left(\chi^s_E\mp \phi^s_E\right)\;,
\end{alignat} 
with\\[-1.cm]
\begin{alignat}{1}
\label{lan34}& \vec{F}_{\pm} =\vec{B}\pm i\vec{E}\;.
\end{alignat} 
The above procedure is equivalent to expressing the KGP equation in the Weyl or chiral representation. This turns out to complicate the understanding of what particle (and antiparticle) solutions are and we will not use this further in our work.

\subsection{Relativistic Covariance} \label{sym}
A well studied ingredient to relativistic quantum mechanics is the symmetry of the equations under Lorentz transformation for boosts and rotations (Poincar\'{e} symmetry by including translations and discrete transformations). More specifically we require that wave equations are a) equivalent when written in another reference frame and b) there exists a prescription of how to relate the equations between the two frames $RF$ and $RF'$. For the Dirac equation this procedure is well-known and will be repeated for completeness. The Dirac equation in the primed frame $RF'$ is
\begin{alignat}{1}
\label{sym01} \left(\gamma'^{\mu}\left(i\hbar c\partial'_{\mu}-eA'_{\mu}\right)-mc^{2}\right)\psi'=0\;.
\end{alignat} 
We note that the gamma matrices will be unchanged during transformation and that four-vectors transform as
\begin{alignat}{1}
\label{sym02} \gamma'^{\mu}=\gamma^{\mu}\;, \qquad A'^{\mu}=a^{\mu}_{\ \nu}A^{\nu}\;,
\end{alignat} 
where $a^{\mu}_{\ \nu}$ are the set of matrices which facilitate the transformations. The primed wave function $\psi'$ must be related to $\psi$ linearly through some matrix $S$
\begin{alignat}{1}
\label{sym03} \psi'=S\psi\;,
\end{alignat} 
which is determined by the structure of Lorentz transformations. For equation Eq.\,\eqref{sym01} to be equivalent to Eq.\,\eqref{kgp02}, the matrix $S$ must satisfy
\begin{alignat}{1}
\label{sym04} S^{-1}\gamma^{\mu}S=a^{\mu}_{\ \nu}\gamma^{\nu}\;.
\end{alignat} 
Using the method of infinitesimal transformations
\begin{alignat}{1}
\label{sym05} a^{\ \nu}_{\mu} = \delta^{\ \nu}_{\mu}+\epsilon^{\ \nu}_{\mu}, \qquad \epsilon_{\mu\nu}=-\epsilon_{\nu\mu}, \qquad S = 1+S_{\mathrm{inf.}}\;,
\end{alignat} 
for arbitrary boosts or rotations $S$ has the form
\begin{alignat}{1}
\label{sym06} S = \mathrm{exp}\left(-\frac{i}{4}\sigma^{\mu\nu}\epsilon_{\mu\nu}\right).
\end{alignat} 
For the KGP equation, the procedure is similar. We will rewrite Eq.\,\eqref{kgp01} as
\begin{subequations}
\begin{alignat}{1}
\label{sym07} &\left(\hbar^{2}c^{2}h_{\mu\nu}\nabla_{A}^{\mu}\nabla_{A}^{\nu}+m^{2}c^{4}\right)\Psi=0\;,\\
&h_{\mu\nu}=g_{\mu\nu}-i\frac{g}{2}\sigma_{\mu\nu}, \qquad \nabla_{A}^{\mu}=\partial^{\mu}+\frac{ie}{\hbar c}A^{\mu}\;,
\end{alignat}
\end{subequations}
where $\nabla_{A}^{\mu}$ is the gauge covariant derivative. The tensor $h_{\mu\nu}$ can be interpreted as an extended metric tensor. In the frame $RF'$, the Eq.\,\eqref{sym07} is
\begin{alignat}{1}
\label{sym08} &\left(\hbar^{2}c^{2}h'_{\mu\nu}\nabla_{A}'^{\mu}\nabla_{A}'^{\nu}+m^{2}c^{4}\right)\Psi'=0\;.
\end{alignat}
As the tensor $h_{\mu\nu}$ is made up of commutators and anti-commutators of $\gamma^{\mu}$, it is unchanged. If we again require that the wave function $\Psi$ transforms linearly then
\begin{alignat}{1}
\label{sym09} h'_{\mu\nu}=h_{\mu\nu}\;, \qquad \Psi'=M\Psi\;.
\end{alignat} 
For the KGP equation to be relativistically covariant then the matrix $M$ must satisfy
\begin{alignat}{1}
\label{sym10} M^{-1}h_{\mu\nu}M=a_{\mu}^{\ \alpha}a_{\nu}^{\ \beta}h_{\alpha\beta}\;.
\end{alignat} 
Since the metric tensor is unchanged by this transformation, the above equation reduces to
\begin{alignat}{1}
\label{sym11} M^{-1}\sigma_{\mu\nu}M=a_{\mu}^{\ \alpha}a_{\nu}^{\ \beta}\sigma_{\alpha\beta}\;.
\end{alignat} 
The g-factor which appears in the tensor $h_{\mu\nu}$ vanishes in Eq.\,\eqref{sym11} which means that g-factor does not effect the Lorentz (or Poincar\'{e}) covariance of the KGP equation. Therefore magnetic moment does not modify Poincar\'{e} symmetry. Without difficulty we can also rewrite Eq.\,\eqref{sym11} using products of Eq.\,\eqref{sym04}, which means that
\begin{alignat}{1}
\label{sym12} S=M\;.
\end{alignat} 
Alternatively this result can be obtained by considering the infinitesimal transformations and making use of the commutator $[\sigma^{\mu\nu},\sigma^{\alpha\beta}]$.
The DP equation also is covariant due to Eq.\,\eqref{sym12}.

\subsection{Conserved currents} \label{currents} 
Before continuing to the Coulomb $1/r$-potential problem which is one of the cases studied in this work we consider how solutions to the KGP equation are to be normalized, asking if the spin structure interferes in some manner. An appropriate Lagrangian that describes the equation of motion in Eq.\,\eqref{kgp01} is given by~\cite{DelgadoAcosta:2010nx}
\begin{alignat}{1}
\label{norm01} &\mathcal{L}=\hbar^{2}c^{2}\left(\nabla_{A}^{\mu\dag}\bar{\Psi}\right)h_{\mu\nu}\left(\nabla_{A}^{\nu}\Psi\right)-m^{2}c^{4}\bar{\Psi}\Psi\;.
\end{alignat}
We note that the field $\Psi$ must have units [length]$^{-1}$, which differs from the Dirac or DP $\psi$ fields, which have units of [length]$^{-3/2}$.

The conserved current can then be expressed as
\begin{alignat}{2}
\label{norm02} \mathcal{J}^{\mu}
=&-\frac{\partial\mathcal{L}}{\partial e\hbar cA_{\mu}}\equiv 
 \mathcal{J}^\mu_{\mathrm{Conv}}+\mathcal{J}^\mu_{\mathrm{Mag}} \\ \notag 
=&i\bar{\Psi}\left(\nabla_{A}^{\mu}\Psi\right)
 -i\left(\nabla_{A}^{\mu\dag}\bar{\Psi}\right)\Psi \\ \notag 
+&\frac{g}{2}\bar{\Psi}\sigma^{\mu}_{\ \beta}\left(\nabla_{A}^{\beta}\Psi\right)
 -\frac{g}{2}\left(\nabla_{A}^{\dag\beta}\bar{\Psi}\right)\sigma_{\beta}^{\ \mu}\Psi\;.
\end{alignat}
The conserved current Eq.\,\eqref{norm02} can be interpreted as the sum of a convection current $\mathcal{J}_{\mathrm{Conv}}$ and magnetization current $\mathcal{J}_{\mathrm{Mag}}$\,. This is nearly identical to the familiar Gordon Decomposition of the Dirac current, with the exception that the magnetization current is proportional to the $g$-factor; the only other difference being that the $\Psi$ and $\psi$ fields have different units and therefore differ in constants in front of the current terms
\begin{subequations}
\begin{alignat}{1}
\label{norm03a}\mathcal{J}^{\mu}_{\mathrm{Conv}}&=i\bar{\Psi}\left(\nabla_{A}^{\mu}\Psi\right)-i\left(\nabla_{A}^{\mu\dag}\bar{\Psi}\right)\Psi,\\
\label{norm03b} 
\mathcal{J}^{\mu}_{\mathrm{Mag}}&=-\frac{g}{2}\partial_{\beta}\left(\bar{\Psi}\sigma^{\beta\mu}\Psi\right)\;.
\end{alignat}
\end{subequations}
The magnetic current Eq.\,\eqref{norm03b} is given by the divergence of the spin density, and by antisymmetry of $\sigma^{\beta\mu}$ it is is conserved. Since the total current $\mathcal{J}^{\mu}$ Eq.\,\eqref{norm02} is conserved, $\mathcal{J}^{\mu}_{\mathrm{Conv}}$ Eq.\,\eqref{norm03a} is also conserved.

\subsection{Physical degrees of freedom}\label{freedom}
The case $g=2$ arises from the product of two covariant Dirac equations differing in sign of $m$, see Eq.\,\eqref{kgp04}. Thus in the set of KGP solutions we obtain contains not four but eight types of basis solutions. Aside of spin ($2\times$) and charge (=particle-antiparticle) ($2\times$), we now also have magnetic moment sign ($2\times$). This is so since in the nonrelativistic reduction the sign of $m$ in the Dirac equation is found in the magnetic moment $\mu\propto g/(\pm m)$. Using the techniques pioneered by Feshbach and Villars~\cite{Feshbach:1958wv}, the KGP equation can be explicitly separated into an 8-component equation~\cite{Robson:1996et,Staudte:1996ey} (these authors restrict themselves to $g=2$) showing the larger space the KGP equation occupies when compared to the Dirac equation.  

This feature of the KGP equation is desirable since it allows us to pick the opposite sign of magnetic moment for particles doublets such as proton and neutron without inventing a new dynamical equation. We do not see a reason why this interpretation of the doubling of the number of degrees of freedom cannot be maintained for $g\ne 2$. 

Like in the Dirac and similarly the KG equations, one must wisely choose the basis states that now belong to a relatively large eight degrees of freedom. We take the convective current as the base for the normalization of eigenstates that in the non-relativistic limit become the solutions of the Pauli equation. The normalization reduces therefore to
\begin{alignat}{1}
\label{norm07}
 N&=2\int dx^{3}\bar{\Psi}_{E}\left(E-eV\right)\Psi_{E}\;.
\end{alignat}
In the KGP case we have second non-positivity originating in the spinor character of the wave function $\Psi_{E}$. Inserting Eq.\,\eqref{Spinor1KGP} into Eq.\,\eqref{norm07} 
\begin{alignat}{1}
N=\!2\!\!\int \!\!d^{3}x\!
&\left(\!\chi_{E}^{\dagger\,s}\left(E-eV\right)\chi_{E}^s
- 
 \phi_{E}^{\dagger\,s}\left(E-eV\right)\phi_{E}^s\right)\;.
\label{norm08} 
\end{alignat}

Unlike the case of the KG equation we are dealing with a spinor equation and we need to set up the meaning of Eq.\,\eqref{norm07} so that particle interpretation is possible. Here we note that KGP normalization combines challenges seen in both Dirac and Klein-Gordon cousins. 

The norm Eq.\,\eqref{norm08} is not positive definite since
\begin{itemize}
\item 
Like in KG equation the coefficient $E-eV$ in Eq.\,\eqref{norm07} can be positive and negative, for the plane wave solutions the norm follows the sign of $E$, and there are both positive and negative energy solutions. In the KG-Boson case one associates this sign with the charge of the Bose-(anti)particle considered.
\item 
In reduction of the spinor Eq.\,\eqref{norm07} to componentwise Eq.\,\eqref{norm08} we recognize the additional sign depending on which component dominates; this is reminiscent of the Dirac equation solution.
\item
A third minus sign arises considering choice of creation of particles and antiparticles sets in the quantum Fermi field operator 
\begin{alignat}{1}
\label{norm07QuantumKGP1} 
\hat \Psi&=\sum_+\hat b_+ \psi_++\sum_-\hat d_-^\dagger \psi_-\;,\\ \notag
 &E_+>0, \qquad E_-<0\;,
\end{alignat}
where like in the Dirac case we assign the negative energy solutions (for weak fields) to be the antiparticle states associated with a creation operator. The sign arises since in bilinear forms we need to commute
\begin{alignat}{1}
\label{norm07QuantumKGP8} 
d_j d_i^\dagger =- d_i^\dagger d_j +\delta_{ij}\;,
\end{alignat}
so that the $d_j|0\rangle=0$ in matrix elements. 
\end{itemize}
All together we have $2^3=8$ different cases corresponding to 8 degrees of freedom of the KGP equation as we noted earlier.

A complete discussion of this matter reaches beyond our objective for a comparative study of solutions of DP and KGP. However, we do want to remark that in order to compensate the additional sign of $E$ in the norm not present in the Dirac equation, we must assign the antiparticle character to the lower two components of the KGP spinor; that is, to the $\phi_{E}^s$ in Eq.\,\eqref{norm08}. The canonical quantization procedure is now implemented as follows: the conjugate momentum $\Pi_\Psi$ to the KGP field $\Psi$ from
\begin{alignat}{1}
\label{QuantumKGP9}
\Pi_\Psi\equiv \frac{\delta I}{\delta \dot \Psi} =\hbar^2 c \nabla_{A}^{\dagger 0}\bar \Psi -\hbar^2 c\frac{g}{2}\vec{\nabla}_{A}^{\dagger}\bar\Psi\cdot\vec{\alpha}\;,
\end{alignat} 
where $I$ is the action and we demand
\begin{alignat}{1}
\label{QuantumKGP10} 
\left\{\Pi_\Psi(\vec x\,\rq,t),\Psi(\vec x,t)\right\}&=i\hbar\delta^3(\vec x-\vec x\,\rq)\;,
\end{alignat}
which requires a full set of all 8 different (real) solutions of KGP equations to form a completeness relation implicit in Eq.\,\eqref{QuantumKGP10}. The charge operator 
\begin{alignat}{1}
\label{QuantumKGP11} 
Q&=-\frac{ie}{2\hbar}\left[\Pi_\Psi(\vec x\,\rq,t),\Psi(\vec x,t)\right] 
\end{alignat}
is organized to count antiparticle states with opposite sign of charge compared to particles. 

\section{Homogeneous magnetic fields} \label{lan}
The simplest situation involving magnetism is the behavior of a particle under the influence of a homogeneous magnetic field
\begin{alignat}{1}
\label{lan01} &\vec{B}=B\hat{z}\;,\end{alignat}
where the magnetic field is chosen to point in the z-direction. Therefore, it is an excellent probe of magnetic moment dynamics. The energy levels of charged particles in such fields are known as Landau levels. In nonrelativistic quantum mechanics with magnetic moment the energy levels are given by
\begin{alignat}{1}
\label{lan02} &E^{NR}=\frac{p_{z}^{2}}{2m}+\frac{e\hbar B}{mc}\left(n+\frac{1}{2}-\frac{g}{2}s\right)\;,\end{alignat}
where the principle quantum number has the values $n=0,1,2\ldots$ and $s=\pm1/2$ describes whether the magnetic moment is aligned or anti-aligned with the magnetic field. For the case of $g=2$, it is common to define the Landau level quantum number
\begin{alignat}{1}
\label{lan03} &\lambda_\mathrm{L}=n+\frac{1}{2}-s\;,\end{alignat}
which has values $\lambda_\mathrm{L}=0,1,2\ldots$. As long as there is no anomalous moment, all the Landau levels except the ground state $\lambda_\mathrm{L}=0$ are double degenerate as there are two combinations of principle quantum number and spin orientation which can produce any given Landau level. 

This degeneracy is broken by the introduction of an anomalous $g\neq2$ moment
\begin{alignat}{1}
\label{lan04} &E^{NR}=\frac{p_{z}^{2}}{2m}+\frac{e\hbar B}{mc}\left(\lambda_\mathrm{L}-as\right)\;.
\end{alignat}
This breaking of degeneracy is shared in the relativistic DP and KGP formulations of this problem, but how that breaking occurs differs between them and will be explored in section~\ref{degener}.

\subsection{The KGP-Landau problem} \label{lankpg}
We see in Eq.\,\eqref{Spinor2KGP} that the problem separates into 2-spinor dynamics.
To solve for the homogeneous magnetic fields in the KGP case we must choose an appropriate gauge. The two most common are the Landau and symmetric gauges,
\begin{alignat}{1}
\label{lan05} \vec{A}_\mathrm{L}=B\left(0,x,0\right)^{T}, \quad \vec{A}_{S}=\frac{B}{2}\left(-y,x,0\right)^{T}\;,
\end{alignat}
respectively. We will choose the symmetric gauge which benefits from preserving rotational symmetry. 

Solving for $\chi$ the KGP, Eq.\,\eqref{kgp01} for homogeneous magnetic fields then becomes
\begin{alignat}{1}
\label{lan06} \Bigg(\left(i\hbar\frac{\partial}{\partial t}\right)^{2}&-\left(i\hbar c\frac{\partial}{\partial x}-e\frac{B}{2}y\right)^{2}-\left(i\hbar c\frac{\partial}{\partial y}+e\frac{B}{2}x\right)^{2}\\ 
\notag &-\left(i\hbar c\frac{\partial}{\partial z}\right)^{2}-m^{2}c^{4}+\frac{g}{2}e\hbar c\vec{\sigma}\cdot\vec{B}\Bigg)\chi=0\;.
\end{alignat}
Considering only energy eigenstates and recognizing that $\vec{p}=-i\hbar\vec{\nabla}$ is the momentum operator and $L_{z}=xp_{y}-yp_{x}$ is the z-component of angular moment, Eq.\,\eqref{lan06} reduces to
\begin{alignat}{1}
\label{lan09} \Big(E^{2}&-m^{2}c^{4}-p^{2}c^{2}+eL_{z}cB\\ \notag &-\frac{1}{4}e^{2}B^{2}\left(x^{2}+y^{2}\right)+\frac{g}{2}e\hbar c\vec{\sigma}\cdot\vec{B}\Big)\chi=0\;.
\end{alignat} 
Using the substitutions 
\begin{alignat}{1}
\label{lan10} E\to m^\prime c^{2}\;,\quad \frac{E^{2}-m^{2}c^{4}}{2E}\to E^\prime\;,
\end{alignat} 
we arrive at the Schroedinger-style Hamiltonian equation
\begin{alignat}{1}
\label{lan11} &E'\chi=\left(\!\frac{p^{2}}{2m^\prime }\!+\!\frac{e^{2}B^{2}}{8m^\prime c^{2}}\left(x^{2}\!+\!y^{2}\right)\!-\!\frac{e}{2m^\prime c}\!\left(\vec{L}+g\vec{S}\right)\!\cdot\! B\hat{z}\!\right)\chi\;,
\end{alignat} 
with the spin operator defined as $\vec{S}=\frac{\hbar}{2}\vec{\sigma}$. 
The Hamiltonian in Eq.\,\eqref{lan11} consists of three separate systems whose operators mutually commute 
\begin{equation}
\label{lan36} H_{Total} =H_\mathrm{HO}+H_\mathrm{free}+H_\mathrm{Mag.}\;.
\end{equation}
We introduce the cyclotron frequency
\begin{alignat}{1}
\label{lan30} &\omega_{c}=eB/m^\prime c\;.
\end{alignat} 
The three terms are:
\begin{subequations}
\begin{itemize}
\item[a)] the harmonic oscillator (HO) with characteristic angular frequency $\omega_{c}=2\omega$
\begin{equation}
\label{lHa} H_\mathrm{HO} =\frac{p_{x}^{2}}{2m^\prime }+\frac{p_{y}^{2}}{2m^\prime }+\frac{1}{2}m^\prime \omega^{2}(x^{2}+y^{2})\;, 
\end{equation}
\item[b)] the free particle Hamiltonian in the z-direction 
\begin{equation}
\label{lHb} H_\mathrm{free} =\frac{p_{z}^{2}}{2m^\prime }\;, 
\end{equation}
and
\item[c)] the magnetic interaction 
\begin{equation}
\label{lHc} H_\mathrm{Mag.}= -\frac{e}{2m^\prime c}(\vec{L}+g\vec{S})\cdot B\hat{z}\;.
 \end{equation}
\end{itemize}
\end{subequations}

\noindent The magnetic interaction Eq.\,\eqref{lHc} is of a familiar form that describes Zeeman splitting including its dependence on the $g$-factor. Because any operator in one Hamiltonian mutually commutes will all operators of the other two Hamiltonians in Eqs.\,\eqref{lHa}, \eqref{lHb}, \eqref{lHc}, e.g $[H,L_{z}]=0$, their energy eigenvalues simply add together. 

\subsection{Ladder operators} \label{ladder}
The Cartesian basis in Eq.\,\eqref{lHa} is undesirable because the $z$-direction angular momentum quantum number $\ell_{z}$ is dependent on the $n_{x}$ and $n_{y}$ HO quantum numbers. It is not too troublesome to disentangle the HO quantum numbers when rewriting the Hamiltonian Eq.\,\eqref{lHa} in terms of ladder operators.

First, however, we move into an auxiliary basis by introducing complex position and momentum variables
\begin{alignat}{1}
\label{lan14} &w=\frac{1}{\sqrt{2}}\left(x-iy\right),\ w^{\dagger}=\frac{1}{\sqrt{2}}\left(x+iy\right)\;,\\ &p_{w}=\frac{1}{\sqrt{2}}\left(p_{x}+ip_{y}\right),\ p_{w}^{\dagger}=\frac{1}{\sqrt{2}}\left(p_{x}-ip_{y}\right)\notag\;,
\end{alignat}
which have the nonzero commutation properties
\begin{alignat}{1}
\label{lan15} &[w,p_{w}]=i\hbar\;.
\end{alignat}
Using the above,the HO Hamiltonian Eq.\,\eqref{lHa} now can be written  
\begin{alignat}{1}
\label{lan16} &H_\mathrm{HO}=\frac{p_{w}^{\dagger}p_{w}}{m^\prime }+m^\prime \omega^{2}w^{\dagger}w\;.
\end{alignat}
We then introduce the ladder operators
\begin{alignat}{1}
\label{lan17} a&=\frac{1}{\sqrt{2}}\left(\beta w+ip_{w}^{\dagger}/\beta \hbar\right),\ b=\frac{1}{\sqrt{2}}\left(\beta w^{\dagger}+ip_{w}/\beta \hbar\right),\\ \notag a^{\dagger}&=\frac{1}{\sqrt{2}}\left(\beta w^{\dagger}-ip_{w}/\beta \hbar\right),\ b^{\dagger}=\frac{1}{\sqrt{2}}\left(\beta w-ip_{w}^{\dagger}/\beta \hbar\right),\\ \notag \beta&=\sqrt{m^\prime \omega/\hbar}\;,\end{alignat}
with nonzero commutation properties
\begin{alignat}{1}
\label{lan18} &[a,a^{\dagger}]=[b,b^{\dagger}]=1\;.
\end{alignat}
This converts Eq.\,\eqref{lan16} into
\begin{alignat}{1}
\label{lan19} &H_\mathrm{HO}=\hbar\omega(a^{\dagger}a+b^{\dagger}b+1)\;.
\end{alignat}
The benefit of all this labor is that $L_{z}$ can be expressed in terms of these ladder operators
\begin{alignat}{1}
\label{lan20} &L_{z}=\hbar(a^{\dagger}a-b^{\dagger}b)\;,\end{alignat}
which transforms the HO Eq.\,\eqref{lHa} and magnetic Eq.\,\eqref{lHc} Hamiltonians into 
\begin{alignat}{1}
\label{lan21} H_\mathrm{HO}&+H_\mathrm{Mag}=\hbar\omega(a^{\dagger}a+b^{\dagger}b+1)\\ \notag &-\hbar\omega(a^{\dagger}a-b^{\dagger}b)-g\omega S_{z} =\hbar\omega(2b^{\dagger}b+1)-g\omega S_{z}\;.
\end{alignat}
This formulation depends only on one pair of ladder operators as determined by the right or left handed nature of the mutually orthogonal $\hat{x}$, $\hat{y}$, and $\hat{B}$ unit vectors; in this case we've chosen the right handed convention. The energy eigenvalues of Eq.\,\eqref{lan21} are
\begin{alignat}{1}
\label{lan22} E_\mathrm{HO}+E_\mathrm{Mag}&=\hbar\omega(2n+1-gs)\\ \notag &=\hbar\omega_{c}(n+\frac{1}{2}-\frac{g}{2}s)\;.
\end{alignat}

The Hamiltonian Eq.\,\eqref{lHa} can be used to determine the operator equations of motion via $d\mathcal{O}/dt=i[H,\mathcal{O}]/\hbar$ yielding the relevant quantities
\begin{subequations}
\begin{alignat}{1}
\label{lan29} m^\prime \frac{dw}{dt}&=i\sqrt{2}\hbar\beta b^{\dagger}=\pi_{w},\\ \frac{d\pi_{w}}{dt}&=-2\omega\sqrt{2}\hbar\beta b^{\dagger}=2i\omega\pi_{w}\;,
\end{alignat}
\end{subequations}
where $\pi_{w}$ is understood to be the kinetic momentum in the complex auxiliary variables. The above two equations combined yield the constant of motion 
\begin{alignat}{1}
\label{lan31} w_{0}=w-\pi_{w}/2im^\prime \omega =x_{0}-iy_{0}\;.
\end{alignat}
This constant of motion corresponds to the center of the particle\rq s orbit and represents an infinite degeneracy of the system in the xy-plane~\cite{Johnson:1950zz}.

The wave function $\chi$ is then a two spinor $\mathrm{X}_{s}$ of the form
\begin{alignat}{1}
\label{lan28} \chi&=\mathrm{X}_{s}\frac{1}{\sqrt{n!}}(a^{\dagger})^{n}\chi_{0}\;,
\qquad 
S_{z}\mathrm{X}_{s}=s\hbar\mathrm{X}_{s}\;,\\ \notag \chi_{0}&=(\beta/\pi)^{1/4}f(w)\mathrm{exp}\Big[ip_{z}z/\hbar-\frac{1}{2}\beta^{2}w^{\dagger}w\Big]\;,\end{alignat}
where the ground state wave function $\chi_{0}$ was arrived at by solving $b\chi_{0}=0$ and $f(w)$ is an arbitrary function that depends on the states $\chi$ being an eigenstate of some combination of Eq.\,\eqref{lan31}. The normalization of Eq.\,\eqref{lan28} depends on the form of $|f(w)|^{2}$, Eq.\,\eqref{norm08}, and the sign of the energy. Now we can finally return to the full Hamiltonian Eq.\,\eqref{lan36}; its eigenvalues rewritten using the ladder operator basis Eq.\,\eqref{lan17} are
\begin{alignat}{1}
\label{lan23} E'&=\frac{p_{z}^{2}}{2m^\prime }+\frac{e\hbar B}{m^\prime c}\left(n+\frac{1}{2}-\frac{g}{2}s\right)\;,\end{alignat}
which depends on the principle quantum number with values $n=0,1,2\ldots$ and the already defined spin orientation numbers. The physical relativistic energies can be obtained by undoing the substitutions in Eq.\,\eqref{lan10} yielding from Eq.\,\eqref{lan23}
\begin{subequations}
\begin{alignat}{1}
\label{lan24} E^{2}&=m^{2}c^{4}+p_{z}^{2}c^{2}+2e\hbar cB\left(n+\frac{1}{2}-\frac{g}{2}s\right),\\[0.2cm] \label{lan24b} 
 E&=\pm\sqrt{m^{2}c^{4}+p_{z}^{2}c^{2}+2e\hbar cB\left(n+\frac{1}{2}-\frac{g}{2}s\right)}\;.
\end{alignat}
\end{subequations}
This expression for the relativistic Landau levels is the same~\cite{Johnson:1950zz,Ferrer:2009nq} as found for the Landau levels for the Dirac equation setting $g=2$ in Eq.\,\eqref{lan24b}. 

\subsection{Nonrelativistic energies} \label{lannonrel}
Restricting ourselves to the positive energy spectrum, the nonrelativistic reduction of Eq.\,\eqref{lan24b} can be carried out in the large mass limit yielding
\begin{alignat}{1}
\label{lan27} E&=mc^{2}+\frac{p_{z}^{2}}{2m}+2\mu_{B}B(\lambda_\mathrm{L}-as)-\frac{p_{z}^{4}}{8m^{3}c^{2}}\\ \notag &\!-\!\frac{p_{z}^{2}}{2m}\frac{2\mu_{B}B}{mc^{2}}(\lambda_\mathrm{L}\!-\!as)\!-\!\frac{2\mu_{B}^{2}B^{2}}{mc^{2}}(\lambda_\mathrm{L}\!-\!as)^{2}\!+\!\mathcal{O}(1/m^{5})\;,\end{alignat}
which contains the expected terms such as the nonrelativistic kinetic energy in the z-direction, the first relativistic correction to kinetic energy, the Landau energies, and cross terms that behave like modifications to the mass of the particle.

\subsection{The DP Landau problem}\label{DPBhom}
The Landau levels for the DP equation are known~\cite{Tsai:1972iq}. They are given by
\begin{subequations}
\begin{alignat}{1}
\label{lan25} 
E^{2}_{DP} =&\left(\!\!\sqrt{\displaystyle m^{2}c^{4}\!+\!2e\hbar cB\lambda_\mathrm{L}}-\frac{eB\hbar}{2mc}(g-2)s\!\right)^{2}\!\!\!+p_{z}^{\!\!2}c^{2},\\[0.4cm]
\label{lan25b}
E_{DP} =\pm &\sqrt{\!\left(\!\!\sqrt{\displaystyle m^{2}c^{4}+2e\hbar cB\lambda_\mathrm{L}}\!-\!\frac{eB\hbar}{2mc}(g-2)s\!\right)^{\!\!2}\!\!\!+p_{z}^{2}c^{2}}\;,
\end{alignat}
\end{subequations}
which in our opinion fails Dirac\rq s principle of mathematical beauty when compared to the KGP result Eq.~\eqref{lan24b}. While both Eqs.~\eqref{lan24b} and \eqref{lan25b} have the correct nonrelativistic reduction at the lowest order to Eq.\,\eqref{lan02}, the latter obscures the physical interpretation. The most egregious issue with the DP-Landau levels is that, in a perturbative expansion, it includes cross terms between the $g=2$ magnetic moment and anomalous terms in $a=(g-2)/2$; thus the result does not depend on the particle magnetic moment alone; there is a functional dependence on the magnetic anomaly $a$. The presence of these cross terms implies that above first order the results cannot be given in terms of the full magnetic moment alone. In contrast, for the KGP-Landau levels, Eq.\,\eqref{lan25b}, the entire effect of magnetic moment is contained in a single term. Thus Dirac\rq s beauty principle favors heavily the KGP.
 
\subsection{State degeneracy}\label{degener}
\begin{figure}
 \centering
 \includegraphics[clip, trim=0.0cm 0.0cm 9.0cm 7.0cm,width=\linewidth]{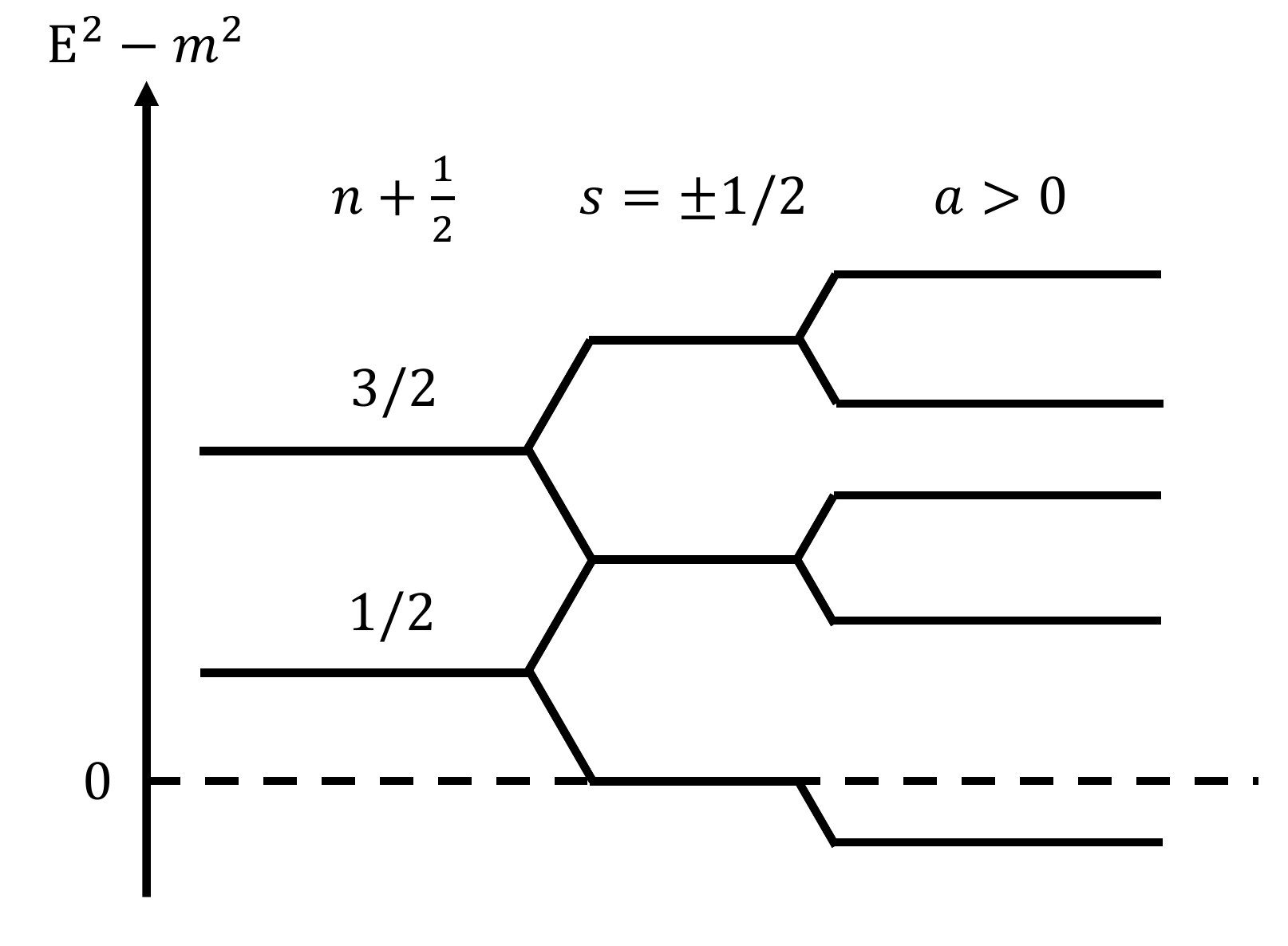}
 \caption[]{Diagram of KGP-Landau levels for particles with zero z-component momentum.}
 \label{f04}
\end{figure}

The KGP-Landau and DP-Landau levels above the ground state lose their (accidental) degeneracy for $g\neq 2$ as we reported below Eq.\,\eqref{lan04}. This is shown schematically in figure~\ref{f04}. The anomaly also causes the ground state to be pushed downward, such that $E^{2}<m^{2}$; if the anomaly and the magnetic field are large enough, states above the ground state are also pushed below the rest mass energy of the particle.

However, in the KGP Eq.\,\eqref{lan24b} we recognize a periodicity considering the energy as a function of $g$. We recall that in Eq.~\eqref{lan24b} $n=0, 1, 2\ldots$. As $g$ varies, each time $gs/2$ crosses an integer value, for a different value of $n$ the energy eigenvalue $E$ repeat as a function of changing $g$. All possible values of energy $E$ are reached (at fixed $m$ and $p^2_z$) for $-2\le g\le 2$. Moreover, while for almost all $g\ne 2$ the degeneracy is completely broken, this periodicity implies that energy degeneracy is restored for values~\cite{Rafelski:2012ui} 
\begin{alignat}{1}
\label{lan26} g_{k}/2=1+k,\ \lambda_\mathrm{L}'=\lambda_\mathrm{L}-ks\;,
\end{alignat}
where $k=0,\pm1,\pm2,\ldots$ The Landau levels Eq.\,\eqref{lan24} contain an infinite number of degenerate levels bounded from below. Certain states change the sign of the magnetic energy and their total energies become unphysical in the limit that $g_{k}B$ becomes large; for even $k$ there are $k/2$ such states and for odd $k$ there are $(k+1)/2$. 

\section{Hydrogen-like atoms} \label{cou}
The hydrogen-like atom provides us a spectroscopic standard candle and one generally explores novel and interesting behavior by comparing against the known Hydrogen-like Dirac spectrum. The Dirac Coulomb solutions are found in many textbooks. The DP case has been discussed extensively by Thaller in Ref.\,\cite{Thaller:1992ji} and we compare with these results. What is missing in literature is a solution of the KGP equation, which we accomplish analytically for the Coulomb potential 
\begin{equation}
\label{cou01} V_\mathrm{C}\equiv e A^{0}=\frac{Z \alpha\hbar c}{r}\;,\qquad \vec{A}=\vec{0}\;.
\end{equation}
Here $\alpha$, when not a vector, denotes the fine structure constant and $Z$ is the integer-charge (in units of $|e|$) of the nucleus. 

\subsection{The KGP-Coulomb problem}
We consider stationary energy states $\Psi=e^{-iEt/\hbar}\Psi_{E}$ and express the derivatives in Eq.\,\eqref{kgp01} in spherical coordinates yielding the differential equation
\begin{alignat}{1}
\label{cou02} \Bigg(\frac{E^{2}-m^{2}c^{4}}{\hbar^{2}c^{2}}&+\frac{Z^{2}\alpha^{2}}{r^{2}}+\frac{2E}{\hbar c}\frac{Z\alpha}{r}+\frac{1}{r}\frac{\partial^{2}}{\partial r^{2}}r\\ \notag &-\frac{L^{2}/\hbar^{2}}{r^{2}}-\frac{g}{2}\frac{e}{2\hbar c}\sigma_{\mu\nu}F^{\mu\nu}\Bigg)\Psi_{E}=0\;.
\end{alignat}
Here we used
\begin{equation}
(-i\hbar\vec{\nabla})^{2}\Psi=-\frac{\hbar^{2}}{r}\frac{\partial^{2}}{\partial r^{2}}(r\Psi)+\frac{L^{2}}{r^{2}}\Psi\;,
\end{equation}
where $L^{2}$ is the orbital angular momentum (squared) operator.

In the laboratory frame the Coulomb field is a pure electrical field without an accompanying magnetic field; therefore the Pauli term only depends on electric field and reduces to
\begin{alignat}{1}
\label{cou03} &-\frac{g}{2}\frac{e}{2\hbar c}\sigma_{\mu\nu}F^{\mu\nu}=-\frac{g}{2}Z\alpha\frac{i\vec{\alpha}\cdot\hat{r}}{r^{2}}\;.
\end{alignat}

It is helpful to introduce Dirac\rq s spin alignment operator
\begin{alignat}{1}
\label{cou04} &\mathcal{K}=\gamma^{0}\left(1+\vec{\Sigma}\cdot\frac{\vec{L}}{\hbar}\right)\;,\end{alignat}
which determines if the spin and orbital angular momentum are aligned or anti-aligned and allows us to write the angular momentum squared operator as
\begin{alignat}{1}
\label{cou05} &L^{2}/\hbar^{2}=\mathcal{K}\left(\mathcal{K}-\gamma^{0}\right)\;.
\end{alignat}
The operator $\mathcal{K}$ commutes with $\vec{\alpha}\cdot\hat{r}$ and its eigenvalues are given as either positive or negative integers $\kappa=\pm(j+1/2)=\pm1,\pm2,\ldots,$ where $j$ is the total angular momentum quantum number. Following the convention of Rose~\cite{r61} the operator $\mathcal{K}$, which share eigenstates of $J^{2}$, are written as
\begin{alignat}{1}
\label{cou18} \mathcal{K}|j,\kappa\rangle =-\kappa|j,\kappa\rangle\;,
\end{alignat}
where $\kappa<0$ are parallel states of $j=\ell+1/2$ while $\kappa>0$ are anti-parallel states of $j=\ell-1/2$ with $\ell\geq1$. 

Equation\,\eqref{cou02} can then be written as
\begin{alignat}{1}
\label{cou06} 0=\Bigg(\!\!\frac{E^{2}\!-\!m^{2}c^{4}}{\hbar^{2}c^{2}}&\!+\!\frac{2E}{\hbar c}\frac{Z\alpha}{r}\!+\!\frac{1}{r}\frac{\partial^{2}}{\partial r^{2}}r\\ \notag &-\frac{\mathcal{K}(\mathcal{K}\!-\!\gamma^{0})\!-\!Z^{2}\alpha^{2}\!-\!\displaystyle\frac{g}{2}Z\alpha(i\vec{\alpha}\cdot\hat{r})}{r^{2}}\Bigg)\Psi_{E}\;.
\end{alignat}
As expected, in relativistic quantum mechanics, the angular momentum eigenvalues must take on non-integer values which in the limit of classical mechanics corresponds to orbits which do not close~\cite{b69}. This \lq\lq effective\rq\rq\ angular momentum depends explicitly on $g$-factor. The difficulty of this equation is that the effective angular momentum operator is non-diagonal in spinor space due to the presence of $\vec{\alpha}\cdot\hat{r}$ which mixes upper and lower components. 

Following the procedure of Martin and Glauber~\cite{Martin:1958zz}, and Biedenharn~\cite{bi62} we introduce the operator
\begin{alignat}{1}
\label{cou07} &\mathfrak{L}=-\gamma^{0}\mathcal{K}-\frac{g}{2}Z\alpha(i\vec{\alpha}\cdot\hat{r})\;,
\end{alignat}
but with the novel modification that $g$-factor directly appears in the second term. This operator commutes with the spin-alignment operator $\mathcal{K}$ and has eigenvalues
\begin{alignat}{1}
\label{cou08} &\Lambda=\pm\sqrt{\kappa^{2}-\displaystyle\frac{\displaystyle g^{2}}{4}Z^{2}\alpha^{2}}\;,\end{alignat}
where the absolute values are denoted as $\lambda=|\Lambda|$. The numerator of the last term in Eq.\,\eqref{cou06} can be then replaced by
\begin{alignat}{1}
\label{cou09} &\mathcal{K}(\mathcal{K}\!-\!\gamma^{0})\!-\!\!Z^{2}\alpha^{2}\!\!-\!\frac{g}{2}Z\alpha(i\vec{\alpha}\cdot\hat{r})\!=\!\mathfrak{L}(\mathfrak{L}\!\!+\!\!1)\!+\!\!\left(\!\!\frac{g^{2}}{4}\!-\!1\!\right)\!\! Z^{2}\alpha^{2}\!.
\end{alignat}

If the $g$-factor is taken to be $g=2$, then the differential Eq.\,\eqref{cou06} reverts to the one discussed in Martin and Glauber\rq s work~\cite{Martin:1958zz}. The coefficient $g^{2}/4-1$ will be commonly seen to precede new more complicated terms, which conveniently vanish for $|g|=2$ demonstrating that as function of $g$ there is a \lq\lq cusp\rq\rq~\cite{Rafelski:2012ui} for $|g|=2$. This will become especially evident when we discuss strongly bound systems in section~\ref{sb}, which behave very differently for $|g|<2$ versus $|g|>2$. 

Eigenstates of both $\mathfrak{L}$ and $E$ then satisfy
\begin{alignat}{1}
\label{cou10} \Bigg(\frac{E^{2}-m^{2}c^{4}}{\hbar^{2}c^{2}}&+\frac{2E}{\hbar c}\frac{Z\alpha}{r}+\frac{1}{r}\frac{\partial^{2}}{\partial r^{2}}r\\ \notag &-\frac{\lambda\left(\lambda\pm1\right)+\left(\displaystyle g^{2}/4-1\right)Z^{2}\alpha^{2}}{r^{2}}\Bigg)\Psi_{E}^{\pm\lambda}=0\;,\end{alignat}
which is diagonal having removed the odd matrix operator in Eq.\,\eqref{cou06} but at a cost of doubling the number of solutions introducing both $\pm\lambda$ spectra. 

It is helpful to introduce dimensionless radial variable $\rho$ and associated quantities
\begin{alignat}{1}
\label{cou11} &\rho=Ar,\quad 
A=2\frac{\displaystyle\sqrt{\displaystyle m^{2}c^{4}-E^{2}}}{\hbar c},\quad 
B=\frac{2Z\alpha E}{\hbar cA}\;.
\end{alignat}
The wave function $\Psi$ is separable into a product of radial and angular functions. We will return to the angular wave functions in section~\ref{ang}. The radial wave function can be substituted with $\Psi\propto U/r$ allowing us to rewrite Eq.\,\eqref{cou10} as
\begin{alignat}{1}
\label{cou12} &\left(\!\frac{\partial^{2}}{\partial\rho^{2}}\!-\!\frac{1}{4}\!+\!\frac{B}{\rho}\!-\!\frac{\lambda(\lambda\pm1)\!+\!\left(\displaystyle g^{2}/4-1\right)\!Z^{2}\alpha^{2}}{\rho^{2}}\right)\! U_{E}^{\pm\lambda}=0\;.
\end{alignat}
The radial wave function, obtained by taking both the $\rho\rightarrow0$ and $\rho\rightarrow\infty$ limits of Eq.\,\eqref{cou12}, is given by
\begin{subequations}
\begin{alignat}{1}
\label{cou14} U_{E}^{\pm\lambda}&=N'\rho^{1/2+\nu}\mathrm{exp}\Big[-\rho/2\Big]F(\rho),\\[0.2cm]
\label{cou14b} \nu&=\sqrt{\left(\lambda\pm1/2\right)^{2}+\left(\frac{g^{2}}{4}-1\right)Z^{2}\alpha^{2}}\;.
\end{alignat}
\end{subequations}

While the general mathematical solution allows both positive and negative $\nu$, only positives values will reproduce the KG-Coulomb or Dirac-Coulomb spectrum for appropriate values of $g$-factor, therefore at this time we will only assume positive values. The other solutions are too singular and cannot be normalized. The radial normalization is then
\begin{alignat}{1}
\label{cou18} 1&=2\int_{0}^{\infty}(E-V_{C})|U_{E}^{\pm\lambda}|^{2}d\rho\;.
\end{alignat}
The orthogonality of the angular parts are considered separately.

The series $F(\rho)$ is the confluent hypergeometric function 
\begin{alignat}{1}
\label{cou16} F(\rho)=1+\frac{a}{c}\rho+\frac{a}{c}\frac{a+1}{c+1}\frac{\rho^{2}}{2}+\ldots,\\
\notag a=\nu+1/2-B,\ c=2\nu+1\;,\end{alignat}
truncated at the term $a+n_{r}=0$ to satisfy Eq.\,\eqref{cou12}, where $n_{r}$ is the node quantum number which takes on the values $n_{r}=0,1,2,\ldots$ The energy levels of the KGP-Coulomb equation are then
\begin{subequations}
\begin{alignat}{1}
\label{cou17} E_{\pm\lambda}^{n_{r},j}&=\frac{mc^{2}}{\sqrt{1+\displaystyle\frac{Z^{2}\alpha^{2}}{\left(n_{r}+1/2+\nu\right)^{2}}}},\\[0.2cm]
\label{cou17b} \nu&=\sqrt{(\lambda\pm1/2)^{2}+\left(\frac{g^{2}}{4}-1\right)Z^{2}\alpha^{2}},\\[0.2cm]
\label{cou17c} \lambda&=\sqrt{\displaystyle(j+1/2)^{2}-\frac{\displaystyle g^{2}}{4}Z^{2}\alpha^{2}}\;.
\end{alignat}
\end{subequations}
Equation\,\eqref{cou17} is the same \lq\lq Sommerfeld-style\rq\rq\ expression for energy that we can obtain from the Dirac or KG equations. The difference between them arises from the expression of the relativistic angular momentum which depends on $g$-factor for the KGP equation. The KGP eigenvalues Eq.\,\eqref{cou17} were also obtained by Niederle and Nikitin~\cite{Niederle:2004bx} using a tensor-spinorial approach for arbitrary half-integer spin particles.

\subsection{Angular functions} \label{ang}
The spherical part of the wave function is given by $\Omega_{\mathfrak{L}}^{\mathcal{K},J_{z}}$, which are orthonormal spherical eigenspinors of the operators $\mathfrak{L}$, $\mathcal{K}$, $J^{2}$, and $J_{z}$. More detailed information for the angular solutions can be found in Martin and Glauber~\cite{Martin:1958zz}. To define the angular part it is first useful to define angular eigenstates of good parity (eigenstates of $\gamma^{0}$), being also eigenstates of $\mathcal{K}$, $J^{2}$, and $J_{z}$ operators
\begin{alignat}{1}
\label{ang01} \gamma^{0}\Omega_{\pm}^{\kappa,j_{z}}=\pm\Omega_{\pm}^{\kappa,j_{z}}\;,\end{alignat}
which can be written as a
\begin{alignat}{1}
\label{ang02} \Omega_{\pm}^{\kappa,j_{z}}&=\\ \notag&\sqrt{\frac{j+j_{z}}{2j}}Y^{j_{z}-1/2}_{j-1/2}\mathrm{X}^{+}_{\pm}+\sqrt{\frac{j-j_{z}}{2j}}Y^{j_{z}+1/2}_{j-1/2}\mathrm{X}^{-}_{\pm}\;,
\end{alignat}
for states where parity and spin orientation eigenvalues have opposite signs $\mathrm{sgn}(\langle\gamma^{0}\rangle)=-\mathrm{sgn}(\langle\mathcal{K}\rangle)$ and
\begin{alignat}{1}
\label{ang03} \Omega_{\pm}^{\kappa,j_{z}}&=\\ \notag&\sqrt{\frac{j-j_{z}+1}{2j+2}}Y^{j_{z}-1/2}_{j+1/2}\mathrm{X}^{+}_{\pm}-\sqrt{\frac{j+j_{z}+1}{2j+2}}Y^{j_{z}+1/2}_{j+1/2}\mathrm{X}^{-}_{\pm}\;,
\end{alignat}
for where parity and spin orientation have the same sign $\mathrm{sgn}(\langle\gamma^{0}\rangle)=\mathrm{sgn}(\langle\mathcal{K}\rangle)$.

The $Y^{m}_{\ell}$ are the traditional orthonormal spherical harmonics and $\mathrm{X}^{\pm}_{\pm}$ are eigenspinors of $\Sigma_{z}$ and $\gamma^{0}$ with
\begin{alignat}{1}
\label{ang04} &\Sigma_{z}\mathrm{X}^{\pm}=\pm\mathrm{X}^{\pm},\ \gamma^{0}\mathrm{X}_{\pm}=\pm\mathrm{X}_{\pm}\;.
\end{alignat}
Because $\mathfrak{L}$ and the parity operator $\gamma^{0}$ do not commute, the states $\Omega_{\mathfrak{L}}^{\mathcal{K},J_{z}}$ are then linear combinations of Eq.\,\eqref{ang02} and Eq.\,\eqref{ang03} given by projecting the good parity states onto states of good $\mathfrak{L}$ via
\begin{alignat}{1}
\label{ang04} &\Omega_{\pm\lambda}^{\kappa<0,j_{z}}=\frac{\lambda\pm\mathfrak{L}}{2\lambda}\Omega_{\mp}^{\kappa<0,j_{z}},\ \Omega_{\pm\lambda}^{\kappa>0,j_{z}}=\frac{\lambda\pm\mathfrak{L}}{2\lambda}\Omega_{\pm}^{\kappa>0,j_{z}}\;.
\end{alignat}
The overall wave function including the radial part Eq.\,\eqref{cou14} is then
\begin{alignat}{1}
\label{ang05} \Psi=N'_{n_{r},\pm\lambda}\rho^{\nu-1/2}\mathrm{exp}\Big[-\rho/2\Big]F(\rho)\Omega_{\pm\lambda}^{\kappa,j_{z}}\;.
\end{alignat}

\subsection{Dirac and Klein-Gordon Spectrum} \label{glimit}
Because we are treating $g$-factor as a unbounded parameter we need to verify that the Dirac and Klein-Gordon energy spectra, which can be found in most texts, see for example Baym~\cite{b69} and Itzykson and Zuber~\cite{iz80}, emerges from the appropriate limit of $g$-factor. In the limit that $g\rightarrow2$ for the Dirac case the expressions for $\lambda$ and $\nu$ reduce to
\begin{subequations}
\begin{alignat}{1}
\label{glimit01} &\lim_{g\rightarrow2}\lambda=\sqrt{\displaystyle(j+1/2)^{2}-Z^{2}\alpha^{2}},\\
&\lim_{g\rightarrow2}\nu_{\pm\lambda}=\lambda\pm1/2\;.
\end{alignat}
\end{subequations}
This procedure requires taking the root of perfect squares; therefore, the sign information is lost in Eq.\,\eqref{glimit01}. As long as $Z^{2}\alpha^{2}<3/4$ we can drop the absolute value notation as $\nu$ is always positive. The energy is then given by
\begin{alignat}{1}
\label{glimit02} &E_{\pm\lambda}^{n_{r},j}=\frac{mc^{2}}{\sqrt{1+\displaystyle\frac{Z^{2}\alpha^{2}}{\left(n_{r}\begin{smallmatrix} +1 \\ +0 \end{smallmatrix}+\sqrt{\displaystyle(j+1/2)^{2}-Z^{2}\alpha^{2}}\right)^{2}}}}\;.
\end{alignat}
The $\begin{smallmatrix} +1 \\ +0 \end{smallmatrix}$ notation is read as the upper value corresponding to the $+\lambda$ states and the lower value corresponding to the $-\lambda$ states. 

The ground state energy (with: $n_{r}=0,\ \Lambda<0,\ j=1/2$) is therefore
\begin{alignat}{1}
\label{glimit07} &E^{0,1/2}_{-\lambda(j=1/2)}=mc^{2}\sqrt{1-Z^{2}\alpha^{2}}\;,\end{alignat}
as expected for the Dirac-Coulomb ground state. Equation\,\eqref{glimit02} reproduces the Dirac-Coulomb energies and also contains a degeneracy between states of opposite $\lambda$ sign, same $j$ quantum number and node quantum numbers offset by one
\begin{alignat}{1}
\label{glimit03} &E^{n_{r}+1,j}_{-\lambda}=E^{n_{r},j}_{+\lambda}\;,\end{alignat}
which will see in section~\ref{nonrel} corresponds to the degeneracy between $2S_{1/2}$ and $2P_{1/2}$ states. There is no degeneracy for the $E^{0,j}_{-\lambda}$ states. 

In the limit that $g\rightarrow 0$, which is the KG case, the expressions are given by
\begin{subequations}
\begin{alignat}{1}
\label{glimit04} &\lim_{g\rightarrow0}\lambda=j+1/2,\\
&\lim_{g\rightarrow0}\nu_{\pm\lambda}=\sqrt{\left(j\begin{smallmatrix} +1 \\ +0 \end{smallmatrix}\right)^{2}-Z^{2}\alpha^{2}}\;,
\end{alignat}
\end{subequations}
which reproduces the correct expressions for the energy levels for the Klein-Gordon case 
\begin{alignat}{1}
\label{glimit05} &E_{\pm\lambda}^{n_{r},j}=\frac{mc^{2}}{\sqrt{1+\displaystyle\frac{Z^{2}\alpha^{2}}{\left(n_{r}+1/2+\displaystyle\sqrt{\left(j\begin{smallmatrix} +1 \\ +0 \end{smallmatrix}\right)^{2}-Z^{2}\alpha^{2}}\right)^{2}}}}\;,\end{alignat}
except that in this limit we are still considering the total angular moment quantum number $j$ rather than orbital momentum quantum number $\ell$. It is interesting to note that the KG-Coulomb problem\rq s energy formula contains $\ell+1/2$, which matches identically to our half-integer $j$ values; therefore, this artifact of spin, untethered and invisible by the lack of magnetic moment, does not alter the energies of the states. The degeneracy in energy levels are given by 
\begin{alignat}{1}
\label{glimit06} &E^{n_{r},j+1}_{-\lambda}=E^{n_{r},j}_{+\lambda}\;,\end{alignat}
with levels of opposite $\lambda$ sign, same node quantum number and shifted $j$ values by one. In a similar fashion to the Dirac case, here we have no degeneracy for $E^{n_{r},1/2}_{-\lambda}$ states. In section~\ref{nonrel} we will convert from $n_{r}$, $j$ and $\pm\lambda$ to the familiar quantum numbers of $n$, $j$ and $\ell$ allowing for easy comparison with the hydrogen spectrum in standard notation.

\subsection{Nonrelativistic limit for energies} \label{nonrel}
The first regime of interest to understand the effect of variable $g$ in the KGP-Coulomb problem is the nonrelativistic limit characterized by the weak binding of low-Z atoms. This will allow us to compare directly with the Schroedinger hydrogen-like atom energies. We start by expanding Eq.\,\eqref{cou17} in powers of $Z\alpha$ to compare to the known hydrogen spectrum. 

To order $\mathcal{O}(Z^{4}\alpha^{4})$ the energy levels are given by
\begin{alignat}{1}
\label{nonrel01} \frac{E^{n_{r},j}_{\pm\lambda}}{mc^{2}}=1&-\frac{1}{2}\frac{Z^{2}\alpha^{2}}{(n_{r}+1/2+(\nu_{\pm\lambda})|_{Z=0})^{2}}\\ 
\notag &+\frac{(\nu_{\pm\lambda})|_{Z=0}'Z^{3}\alpha^{3}}{(n_{r}+1/2+(\nu_{\pm\lambda})|_{Z=0})^{3}}\\ 
\notag&+\frac{1}{2}\frac{(3/4-3(\nu_{\pm\lambda})|_{Z=0}'^{2})Z^{4}\alpha^{4}}{(n_{r}+1/2+(\nu_{\pm\lambda})|_{Z=0})^{4}}\\ 
\notag &+\frac{1}{2}\frac{(\nu_{\pm\lambda})|_{Z=0}^{\prime\prime} Z^{4}\alpha^{4}}{(n_{r}+1/2+(\nu_{\pm\lambda})|_{Z=0})^{3}}+\mathcal{O}(Z^{6}\alpha^{6})\;,\end{alignat}
where primed $\nu_{\pm\lambda}$ indicate derivatives with respect to $Z\alpha$. These derivatives evaluate to
\begin{alignat}{1}
\label{nonrel02} &(\nu_{\pm\lambda})|_{Z=0}=j+1/2\pm1/2,\\ \notag &(\nu_{\pm\lambda})|_{Z=0}'=0,\\ \notag &(\nu_{\pm\lambda})|_{Z=0}^{\prime\prime}=\frac{(g^{2}/4-1)}{j+1/2\pm1/2}-\frac{g^{2}/4}{j+1/2}\;.
\end{alignat}
Equation\,\eqref{nonrel01} then simplifies to
\begin{alignat}{1}
\label{nonrel03} \frac{E^{n_{r},j}_{\pm\lambda}}{mc^{2}}=1&-\frac{1}{2}\frac{Z^{2}\alpha^{2}}{\left(n_{r}+j\begin{smallmatrix}+3/2 \\ +1/2\end{smallmatrix}\right)^{2}}\\
\notag &+\frac{3}{8}\frac{Z^{4}\alpha^{4}}{\left(n_{r}+j\begin{smallmatrix}+3/2 \\ +1/2\end{smallmatrix}\right)^{4}}\\ 
\notag &+\frac{1}{2}\left(\frac{(g^{2}/4-1)}{j\begin{smallmatrix}+1 \\ +0\end{smallmatrix}}-\frac{g^{2}/4}{j+1/2}\right)\frac{Z^{4}\alpha^{4}}{\left(n_{r}+j\begin{smallmatrix}+3/2 \\ +1/2\end{smallmatrix}\right)^{3}}\\
\notag &+\mathcal{O}(Z^{6}\alpha^{6})\;.
\end{alignat}
In the non relativistic limit, the node quantum number corresponds to the principle quantum number via $n_{r}=n'-j-1/2$ with $n'=1,2,3\ldots$ Using Eqs.\,\eqref{nonrel03} and \eqref{cou07} we see that in the non relativistic limit $+\lambda$ corresponds to $\kappa>0$ or anti-aligned spin-angular momentum with $j=\ell-1/2$ and $\ell\geq1$. Conversely $-\lambda$ corresponds to $\kappa<0$ or aligned spin-angular momentum with $j=\ell+1/2$. 

With all this input we arrive at
\begin{alignat}{1}
\label{nonrel04} \frac{E^{n,j}_{\begin{smallmatrix}
\kappa>0 \\ \kappa<0
\end{smallmatrix}}}{mc^{2}}=1&-\frac{1}{2}\frac{Z^{2}\alpha^{2}}{\left(n'\begin{smallmatrix}
+1 \\ +0
\end{smallmatrix}\right)^{2}}+\frac{3}{8}\frac{Z^{4}\alpha^{4}}{\left(n'\begin{smallmatrix}
+1 \\ +0
\end{smallmatrix}\right)^{4}}
\\ \notag &+\frac{1}{2}
\frac{(g^{2}/4-1)}{j\begin{smallmatrix}
+1 \\ +0
\end{smallmatrix}}\frac{Z^{4}\alpha^{4}}{\left(n'\begin{smallmatrix}
+1 \\ +0
\end{smallmatrix}\right)^{3}}\\
\notag &-\frac{1}{2}\frac{g^{2}/4}{j+1/2}\frac{Z^{4}\alpha^{4}}{\left(n'\begin{smallmatrix}
+1 \\ +0
\end{smallmatrix}\right)^{3}}+\mathcal{O}(Z^{6}\alpha^{6})\;.
\end{alignat}
Lastly we recast, for the $\kappa>0$ states, the principle quantum number as $n'+1\rightarrow n$ with $n\geq2$ and we simply relabel $n'\rightarrow n$ for $\kappa<0$ states. This allows Eq.\,\eqref{nonrel04} to be completely written in terms of $n$, $j$, and $\ell$ as
\begin{alignat}{1}
\label{nonrel09} \frac{E^{n,j}_{\ell}}{mc^{2}}=1&-\frac{1}{2}\frac{Z^{2}\alpha^{2}}{n^{2}}+\frac{3}{8}\frac{Z^{4}\alpha^{4}}{n^{4}}\\
\notag &+\frac{1}{2}\frac{(g^{2}/4-1)}{\ell+1/2}\frac{Z^{4}\alpha^{4}}{n^{3}}\\
\notag &-\frac{1}{2}\frac{g^{2}/4}{j+1/2}\frac{Z^{4}\alpha^{4}}{n^{3}}+\mathcal{O}(Z^{6}\alpha^{6})\;,\end{alignat}
where it is understood that $n-\ell\geq1$, this condition allows us to write what was previously described in Eq.\,\eqref{nonrel04} as two distinct spectra now as a single energy spectra. In the limit $g\rightarrow2$ or $g\rightarrow0$ the correct expansion to order $Z^{4}\alpha^4$ of the Dirac or KG energies are obtained. In the following we explore some consequences of our principal nonrelativistic result, Eq.\,\eqref{nonrel09}.

\subsection{Lamb Shift} \label{lamb}
The breaking of degeneracy in Eq.\,\eqref{nonrel09} between states of differing $\ell$ orbital quantum number, but the same total angular momentum $j$ and principle quantum number $n$ is responsible for the Lamb shift due to anomalous magnetic moment. The only term in Eq.\,\eqref{nonrel09} (up to order $Z^{4}\alpha^{4}$) that breaks the degeneracy between the $E^{n,j}_{\ell=j+1/2}$ and $E^{n,j}_{\ell=j-1/2}$ states for $n\geq2$ is the fourth term. This is unsurprising as it depends exclusively on quantum number $\ell$ and $n$. The lowest order Lamb shift due to anomalous magnetic moment is then
\begin{alignat}{1}
\label{lamb01} \frac{\Delta E_{\mathrm{gLamb}}^{n,j}}{mc^{2}}&=E^{n,j}_{\ell=j-1/2}-E^{n,j}_{\ell=j+1/2}\\ \notag &=\left(g^{2}/8-1/2\right)\left(\frac{1}{j}-\frac{1}{j+1}\right)\frac{Z^{4}\alpha^{4}}{n^{3}}\;.
\end{alignat}
For the $2S_{1/2}$ and $2P_{1/2}$ states Eq.\,\eqref{lamb01} reduces to
\begin{alignat}{1}
\label{lamb02} \frac{\Delta E_{\mathrm{gLamb}}^{2S_{1/2}-2P_{1/2}}}{mc^{2}}&=\left(g^{2}/8-1/2\right)\frac{Z^{4}\alpha^{4}}{6}\\ \notag &=\left(a+a^{2}/2\right)\frac{Z^{4}\alpha^{4}}{6}\;.
\end{alignat}
Our result in Eqs.\,\eqref{lamb01} and \eqref{lamb02} is sensitive to $g^{2}/8-1/2=a+a^{2}/2$. Traditionally the Lamb shift due to an anomalous lepton magnetic moment is obtained perturbatively~\cite{iz80} by considering the DP equation which is sensitive to $g/2-1=a$ the shift takes on the expression at lowest order
\begin{alignat}{1}
\label{lamb03} \frac{\Delta E_{\mathrm{gLamb,DP}}^{2S_{1/2}-2P_{1/2}}}{mc^{2}}&=\left(\frac{g-2}{2}\right)\frac{Z^{4}\alpha^{4}}{6}\\ \notag &=a\frac{Z^{4}\alpha^{4}}{6}\;.
\end{alignat}
It is of experimental interest to resolve this discrepancy between the first order DP equation and the second order fermion formulation KGP. We recall the present day values
\begin{subequations}
\begin{alignat}{1}
\label{aeFULL} a_e&=1159.65218091(26)\times 10^{-6}\simeq \frac{\alpha}{2\pi}\;,\\
a_\mu-a_e&=6.2687(6)\times 10^{-6}\;.
\end{alignat}
\end{subequations}
The largest contribution to the anomalous moment for charged leptons is, as indicated the lowest order QED Schwinger result $a=\alpha/2\pi $. For the KGP approach, the anomalous $g$-factor mixes contributions of different powers of fine structure $\alpha$. Precision values for the fundamental constants are taken from~\cite{Mohr:2012tt}. For the $2S_{1/2}–2P_{1/2}$ states, the shift is
\begin{alignat}{1}
\label{lamb04} &\frac{\Delta E_{\mathrm{gLamb}}^{2S_{1/2}–2P_{1/2}}}{mc^{2}}=\frac{Z^{4}\alpha^{5}}{12\pi}+\frac{Z^{4}\alpha^{6}}{48\pi^{2}}\;.
\end{alignat}
The scale of the discrepancy between KGP and DP for the hydrogen atom is then
\begin{alignat}{1}
\label{lamb05} \Delta &E_{\mathrm{gLamb,KGP}}^{2S_{1/2}-2P_{1/2}}\!-\!\Delta E_{\mathrm{gLamb,DP}}^{2S_{1/2}-2P_{1/2}}\!=\!\frac{\alpha^{6}mc^{2}}{48\pi^{2}}\\ \notag&\;\;=\!1.62881214\times10^{-10}\;\mathrm{eV} =39.3845030\;\mathrm{kHz}\;,\end{alignat}
without taking into account the standard corrections such as reduced mass, recoil, radiative, or finite nuclear size; for more information on those corrections please refer to~\cite{sy90,Jentschura:1996zz,Eides:2000xc,Mohr:2012tt}. It is to be understood that the corrections presented here are illustrative of the effect magnetic moment has on the spectroscopic levels, but that further work is required to compare these to experiment: for example we look here on behavior of point particles only.

While the discrepancy is small for the hydrogen system, it is $\approx 40$ kHz and will be visible in this or next generation\rq s spectroscopic experiments. The discrepancy is also non-negligible for hydrogen-like exotics such as proton-antiproton because the proton $g$-factor is much larger
\begin{alignat}{1}\label{gpaFULL}
&g_p=5.585694702(17)\;,\qquad a_p=1.792847351(9)\;. 
\end{alignat} 
The discrepancy for the proton-antiproton system is
\begin{alignat}{1}
\label{lamb06} \Delta E_{\mathrm{gLamb,KGP}}^{2S_{1/2}-2P_{1/2}}-\Delta E_{\mathrm{gLamb,DP}}^{2S_{1/2}-2P_{1/2}}&=0.71268151\;\mathrm{eV}\;.
\end{alignat} 
The Lamb shift is of interest to the muonic-hydrogen system as is used to measure the proton charge radius $R_p^{\mu\mathrm{LS}}$. Experimentally, the exotic atom study yields~\cite{Pohl:2013yb}
\begin{alignat}{1}\label{protonR}
R_p^{\mu\mathrm{LS}}=0.84087(39)\;\mathrm{fm}\;,
\end{alignat} 
which is 4\% off the CODATA value for the proton charge radius $R_p^\mathrm{COD}$
\begin{equation}
R_p^\mathrm{COD}=0.8775(51)\;\mathrm{fm}\;,
\end{equation}
a discrepancy of 7 sigma. In terms of the Lamb shift, assuming the CODATA value, this corresponds to an unexplained additional shift of approximately $0.31$\;meV. The modification to the $\mu P$ Lamb shift due to the KGP equation for muonic-hydrogen, the KGP-DP discrepancy is
\begin{alignat}{1}
\label{lamb07} \Delta E_{\mathrm{gLamb,KGP}}^{2S_{1/2}-2P_{1/2}}-\Delta E_{\mathrm{gLamb,DP}}^{2S_{1/2}-2P_{1/2}}&=\!3.394064\times10^{-8}\;\mathrm{eV}\;,
\end{alignat} 
which is too small to explain the result of Pohl \emph{et. al.} \cite{Pohl:2013yb}.

However, this does not disqualify the possible magnetic moment involvement in this discrepancy: the magnetic moment coupling via KGP equation should be considered in the study of scattering cross section of electrons on protons which is a decisive input into the CODATA proton radius value. Moreover, we see effects related to other relativistic forms of magnetic moment interaction which can modify the muonic Lamb shift. We return to these questions at the end of this work.

\subsection{Fine structure} \label{fs}
The fifth term in Eq.\,\eqref{nonrel09}, which depends on $j$ and $n$, will shift the levels due to an anomalous moment, but does not contribute to the Lamb shift. Rather this expression, which contains the spin-orbit $\vec{L}\cdot\vec{S}$ coupling, is responsible for the fine structure splittings. From Eq.\,\eqref{nonrel09} the fine structure splitting is given by
\begin{alignat}{1}
\label{fs00} \frac{\Delta E_{\mathrm{gFS}}^{n,\ell}}{mc^{2}}&=E^{n,j=\ell+1/2}_{\ell}-E^{n,j=\ell-1/2}_{\ell}\\ \notag &=\left(g^{2}/8\right)\left(\frac{1}{\ell}-\frac{1}{\ell+1}\right)\frac{Z^{4}\alpha^{4}}{n^{3}}\;.
\end{alignat}
The splitting between the $2P_{3/2}$ and $2P_{1/2}$ states is therefore
\begin{alignat}{1}
\label{fs01} \frac{\Delta E_{\mathrm{gFS}}^{2P_{3/2}-2P_{1/2}}}{mc^{2}}&=\left(g^{2}/8\right)\frac{Z^{4}\alpha^{4}}{16}\\ \notag\ &=\left(1/2+a+a^{2}/2\right)\frac{Z^{4}\alpha^{4}}{16}\;.
\end{alignat}
In comparison the fine structure dependence on $g$-factor in the DP equation is given as
\begin{alignat}{1}
\label{fs02} \frac{\Delta E_{\mathrm{gFS,DP}}^{2P_{3/2}-2P_{1/2}}}{mc^{2}}&=\left(\frac{g-1}{2}\right)\frac{Z^{4}\alpha^{4}}{16}\\ \notag\ &=\left(1/2+a\right)\frac{Z^{4}\alpha^{4}}{16}\;.
\end{alignat}
Just as in the case of the Lamb shift, we find that the KGP and DP equations disagree for fine structure splitting. For the hydrogen atom this discrepancy is 
\begin{alignat}{1}
\label{fs03} \Delta &E_{\mathrm{gFS,KGP}}^{2P_{3/2}-2P_{1/2}}\!-\!\Delta E_{\mathrm{gFS,DP}}^{2P_{3/2}-2P_{1/2}}\!=\!\frac{\alpha^{6}mc^{2}}{128\pi^{2}}\\ \notag&\;\;=\!6.10804553\times10^{-11}\;\mathrm{eV}=14.7691885\;\mathrm{kHz}\;,\end{alignat}
and for proton-antiproton, the fine structure splitting discrepancy is
\begin{alignat}{1}
\label{fs04} \Delta E_{\mathrm{gFS,KGP}}^{2P_{3/2}-2P_{1/2}}-\Delta E_{\mathrm{gFS,DP}}^{2P_{3/2}-2P_{1/2}}&=0.26725557\;\mathrm{eV}\;.
\end{alignat}
For fine structure of the muonic-hydrogen system, the KGP-DP discrepancy is
\begin{alignat}{1}
\label{fs05} \Delta E_{\mathrm{gFS,KGP}}^{2S_{1/2}-2P_{1/2}}-\Delta E_{\mathrm{gFS,DP}}^{2S_{1/2}-2P_{1/2}}&=\!1.272774\times10^{-8}\;\mathrm{eV}\;
\end{alignat} 
We can make a general observation that non minimal magnetic coupling, such as we have studied in the DP and KGP cases, enlarge energy level splittings. The above shows that these discrepancies will remain when calculating within more realistic finite nuclear size context.

\section{Critical binding} \label{sb}
\subsection{Applicability of DP or KGP to critical fields}\label{vacfl}
Care must be taken when interpreting the results presented in section~\ref{lan}. For physical electrons the AMM interaction is the result of vacuum fluctuations whose strength also depends on the strength of the field. For example in the large magnetic field limit a QED computation shows that the ground state is instead of Eq.~\eqref{lan25b} given by~\cite{Jancovici:1970ep}
\begin{alignat}{1} \label{vacfl01}
E_{0}\approx mc^{2}+\frac{\alpha}{4\pi}mc^{2}\ \mathrm{ln}^{2}\left(\frac{2e\hbar B}{m^{2}c^{3}}\right)
\end{alignat}
which even for enormous magnetic fields does not deviate significantly from the rest mass-energy of the electron. Further the AMM  radiative corrections approach zero for higher Landau levels~\cite{Ferrer:2015wca}. Therefore the AMM in the case of electrons does not have a significant effect in highly magnetized environments such as those found in astrophysics (magnetars).

The situation is different for composite particles such as the proton, neutron and light nuclei whose anomalous magnetic moments are dominated by their internal structure and not by vacuum fluctuations. In this situation we expect that the AMM interaction in high magnetic fields remains significant. Therefore, asking whether the DP or KGP equations better describes the dynamics of composite hadrons and atomic nuclei in presence of magnetar strength fields is a relevant question despite the standard choice in literature being the DP equation~\cite{Broderick:2000pe}. The same question can be asked for certain exotic hydrogen-like atoms where the constituent particles have anomalous moments which can be characterized as an external parameter. 

\subsection{Homogeneous magnetic fields} \label{sbl}
\begin{figure}
 \centering
 \includegraphics[clip, trim=0.0cm 0.0cm 0.0cm 0.5cm,width=\linewidth]{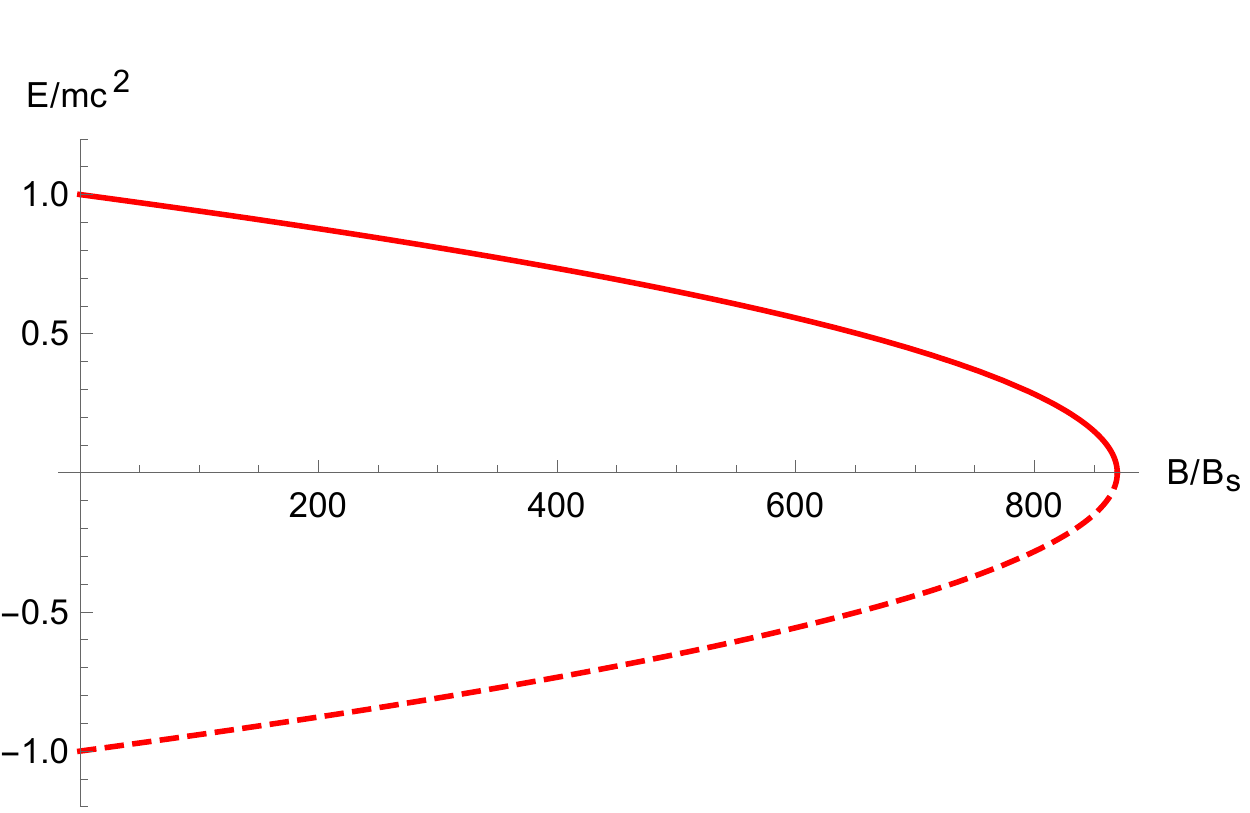}
 \caption[]{The $n=0$, $s=1/2$ ground state for a KGP electron given by Eq.~\eqref{lan24b} with $g/2-1=\alpha/2\pi$ in a homogeneous magnetic field. We consider the particle with no $z$-direction momentum. The particle state (solid red) and antiparticle (dashed red) are presented.}
 \label{f01}
\end{figure}

As noted in section~\ref{ladder} the magnetic moment anomaly can flip the sign of the magnetic energy for the least excited states causing the gap between particle and antiparticle states to decrease with magnetic field strength. Setting $p_z=0$ in Eq.\,\eqref{lan24}, we show in figure~\ref{f01} that the energy of the lowest KGP Landau eigenstate $n=0, s=1/2$ reaches zero where the gap between particle and antiparticle states vanishes for the field
\begin{subequations}
\begin{alignat}{1}\label{Bcrit}
&B_\mathrm{crit}^{e}=\frac{\mathcal{B}_{S}^{e}}{a_{e}}=861\mathcal{B}_{S}^{e} =3.8006\times10^{16}\;\mathrm{G}\;,\\
&B_\mathrm{crit}^{p}=\frac{\mathcal{B}_{S}^{p}}{a_{p}}=\frac{1}{1.79}\mathcal{B}_{S}^{p}=8.3138\times10^{19}\;\mathrm{G}\;,
\end{alignat} 
\end{subequations}
where $\mathcal{B}_{S}$ is the so-called Schwinger critical field~\cite{Schwinger:1951nm}.
\begin{subequations}
\begin{alignat}{1}\label{Bsch}
\mathcal{B}_{S}^{e}\equiv\frac{{m_{e}^2}c^3}{e\hbar}=\frac{m_{e}c^2}{2\mu_B}=4.4141\times 10^{13}\;\mathrm{G}\;,\\
\mathcal{B}_{S}^{p}\equiv\frac{{m_{p}^2}c^3}{e\hbar}=\frac{m_{p}c^2}{2\mu_N}=1.4882\times 10^{20}\;\mathrm{G}\;.
\end{alignat}
\end{subequations}
The numerical results are evaluated for the anomalous moment of the electron and proton, given by Eq.\eqref{aeFULL} and \eqref{gpaFULL}. At the critical field strength $B_\mathrm{crit}$ the Hamiltonian loses self-adjointness and the KGP loses its predictive properties. The Schwinger critical field Eq.\,\eqref{Bsch} denotes the boundary when electrodynamics is expected to behave in an intrinsically nonlinear fashion, and the equivalent electric field configurations become unstable~\cite{Labun:2008re}. However, it is also possible that the vacuum is stabilized by such strong magnetic fields~\cite{Evans:2018kor}.
 
The critical magnetic fields as shown in Eq.\,\eqref{Bcrit} appear in discussion of magnetars~\cite{Kaspi:2017fwg}. The magnetar field is expected to be more than 100-fold that of the Schwinger critical magnetic field which is on the same order of magnitude as $B_\textrm{crit}$ for an electron. While the critical field for a proton exceeds that of a magnetar, the dynamics of protons (and neutrons) in such fields is nevertheless significantly modified. A correct description of magnetic moment therefore has relevant consequences to astrophysics. 

\begin{figure}
 \centering
 \includegraphics[clip, trim=0.0cm 0.0cm 0.0cm 0.5cm,width=\linewidth]{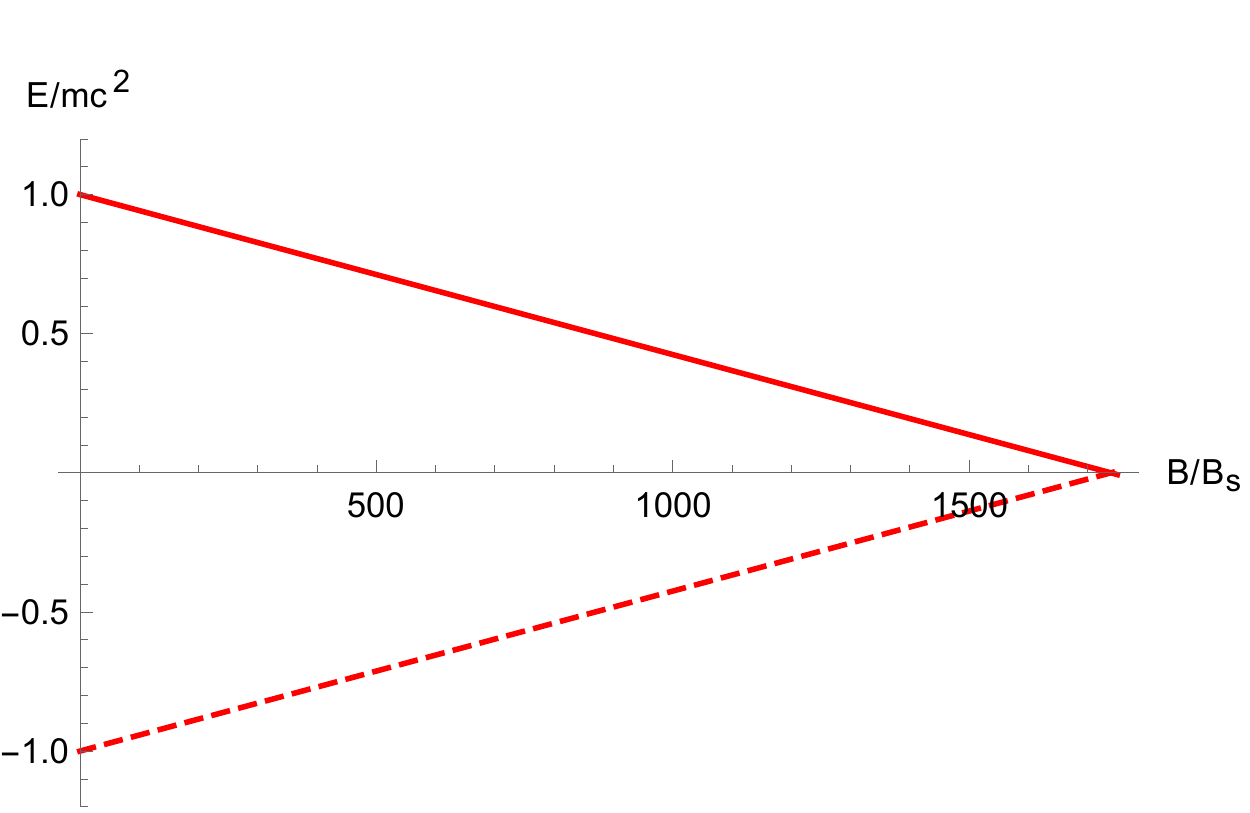}
 \caption[]{The $n=0$, $s=1/2$ ground state for a DP electron given by Eq.~\eqref{lan25b} with $g/2-1=\alpha/2\pi$ in a homogeneous magnetic field. We consider the particle with no $z$-direction momentum. The particle state (solid red) and antiparticle (dashed red) are presented.}
 \label{f02}
\end{figure}

Figure~\ref{f02} shows analogous reduction in particle/anti\-particle energy gap for the DP equation. In this case the vanishing point happens at a larger magnetic field strength. This time the solutions continue past this point, but require allowing the states to cross into the opposite continua which we consider unphysical. We are not satisfied with either model\rq s behavior though the KGP description is preferable for reasons stated earlier in section~\ref{lan}. However, it is undesirable that both KGP and DP solutions loose physical meaning and vacuum stability in strong magnetic fields.

\subsection{Extension to the KGP Equation}\label{IKGP}
We have seen that at a sufficiently strong magnetic field an unexpected instability can occur in the presence of magnetic field alone. Of interest is a further extension capable of restoring the stability of the system. We show that such an improvement of the KGP equation is possible. This opens a new research direction which is beyond the scope of this work. However, it is so easy to show how this works that we cannot resist the temptation. The approach is based on noting a self-evident relationship between magnetic moment and mass. 

The constituent equation, which we will refer to as the Improved-KGP or IKGP, takes the form
\begin{alignat}{1}
\label{IKGP01} &\left(\left(i\hbar c\partial_{\mu}-eA_{\mu}\right)^{2}-\left(mc^{2}+\frac{g}{4}\mu_{B}\sigma^{\mu\nu}F_{\mu\nu}\right)^{2}\right)\Psi=0\;.
\end{alignat} 
We introduce 
\begin{alignat}{1}
\label{IKGP02} &\widetilde{m}=m+\frac{g}{4}\frac{\mu_{B}}{c^{2}}\sigma^{\mu\nu}F_{\mu\nu}\;,
\end{alignat} 
as an effective magnetic mass which is off-diagonal in spinor-space in the Dirac representation. IKGP differs from the KGP by the presence of the additional interaction term
\begin{alignat}{1} \label{IKGP03}
\delta V\equiv &\left(\frac{g}{2}\mu_{B}\right)^{2}\left(\frac{1}{2}\sigma^{\mu\nu}F_{\mu\nu}\right)^{2}\;,
\end{alignat}
which is proportional to the square of the magnetic moment 
\begin{alignat}{1}
\label{IKGP04} &\left(\frac{1}{2}\sigma^{\mu\nu}F_{\mu\nu}\right)^{2}=2\left(\mathcal{S}+\gamma^{5}\mathcal{P}\right)\;,
\end{alignat}
where the invariants of the EM field are defined as
\begin{alignat}{1}
&\mathcal{S}=\frac{1}{2}(B^{2}-E^{2}), \qquad \mathcal{P}=\vec{E}\cdot\vec{B}\;.
\end{alignat}
This modification can be thought of as \lq\lq completing the square\rq\rq of KGP. 

For the homogeneous magnetic field the IKGP equation can be solved in much the same way as the KGP equation in section~\ref{lan}. One obtains eigenvalues noting a simple shift that occurs in view of Eq.~\eqref{IKGP03} which predicts a shift of $m^2$ quadratic in the magnetic field. The resulting energy levels are
\begin{alignat}{1}
\label{IKGP05} E&=\pm\sqrt{m^{2}c^{4}+\left(\frac{g}{2}\mu_{B}\right)^{2}B^{2}+2e\hbar cB\left(n+\frac{1}{2}-\frac{g}{2}s\right)}\;.
\end{alignat}
An interesting feature is that in the ultra-high magnetic fields ($B>>\mathcal{B}_{s}$), Eq.~\eqref{IKGP05} approximates
\begin{alignat}{1}
\label{IKGP07} E\approx\frac{g}{2}\mu_{B}B\;,
\end{alignat}
which is not dissimilar to the non-relativistic case where the magnetic energy is simply proportional to the magnetic field.

\begin{figure}
 \centering
 \includegraphics[clip, trim=0.0cm 0.0cm 0.0cm 0.5cm,width=\linewidth]{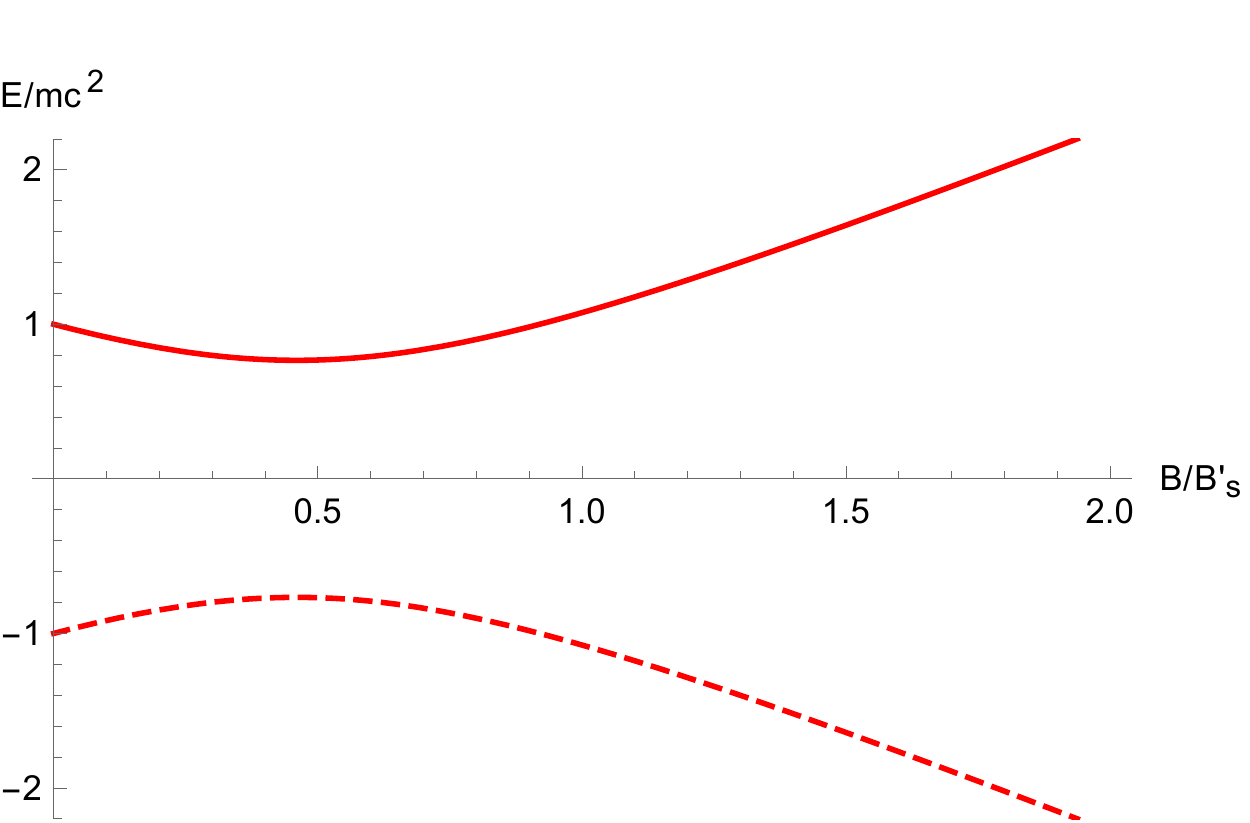}
 \caption[]{The $n=0$, $s=1/2$ ground state for a IKGP proton given by Eq.~\eqref{IKGP05} with $g=5.58$ in a homogeneous magnetic field. The magnetic minimum is well visible for particles with larger anomalous moment such as proton. We consider the particle with no $z$-direction momentum. The particle state (solid red) and antiparticle (dashed red) are presented. The magnetic field scale is $\mathcal{B}'_{S}=(m^2_p/m^2_e)\mathcal{B}_S$.}
\label{f05}
\end{figure}

The most striking feature is that the ground state remains physical for all values of magnetic field when an anomalous moment is included and the self-adjointness of the system is not lost for some critical magnetic field strength. It can be then thought that the magnetic field provides a stabilizing influence on the system. Rather, there exists a \lq\lq magnetic minimum\rq\rq\ located for $n=0, s=1/2$ at
\begin{alignat}{1}
\label{IKGP06} B_{\mathrm{min}}&=\frac{4mc^{2}}{g^{2}\mu_{B}}a\;,
\end{alignat}
which for an electron is
\begin{alignat}{1}
B_{\mathrm{min}}^{e}&=\frac{8a_e}{g_e^2}\mathcal{B}_S
=1.02126\times10^{11}\;\mathrm{G}\;.
\end{alignat}
We are in particular interested in the environment of the magnetar stars. We thus evaluate using Eq.~\eqref{gpaFULL} the minimum for a proton
\begin{alignat}{1}
B_{\mathrm{min}}^{p}&=\frac{8a_p}{g_p^2} \frac{m_p^2}{m_e^2}\mathcal{B}_s
=6.841\times10^{19}\;\mathrm{G}\;,
\end{alignat}
which can be seen in figure~\ref{f05}. Here it is understood that for the calculation of the proton's magnetic minimum, the nuclear mass and magneton was used rather than the electron Bohr magneton. For large enough g-factor, excited states may also contain a minimum, but for any nonzero anomalous moment the ground state always does.

\subsection{The Coulomb problem} \label{sbc}
For the case of $g=2$ hydrogen-like systems with large $Z$ nuclei, there is extensive background related to the long study of the solutions of the Dirac equation~\cite{Rafelski:1976ts,Greiner:1985ce,Rafelski:2016ixr}. For $g\ne 2$ and $1/r$ singular potential we refer back to the exact expression for the energy levels in Eq.\,\eqref{cou17}. In the situation of critical electric fields, states lose self-adjointness for large $Z$. For $|g|<2$, just as in the Dirac energy levels for $1/r$ singular potential, but if $|g|>2$ there is merging of particle to particle states (and antiparticle to antiparticle) for states of the same total angular momentum quantum number $j$, but opposite spin orientations.

\begin{figure}
 \centering
 \includegraphics[width=\linewidth]{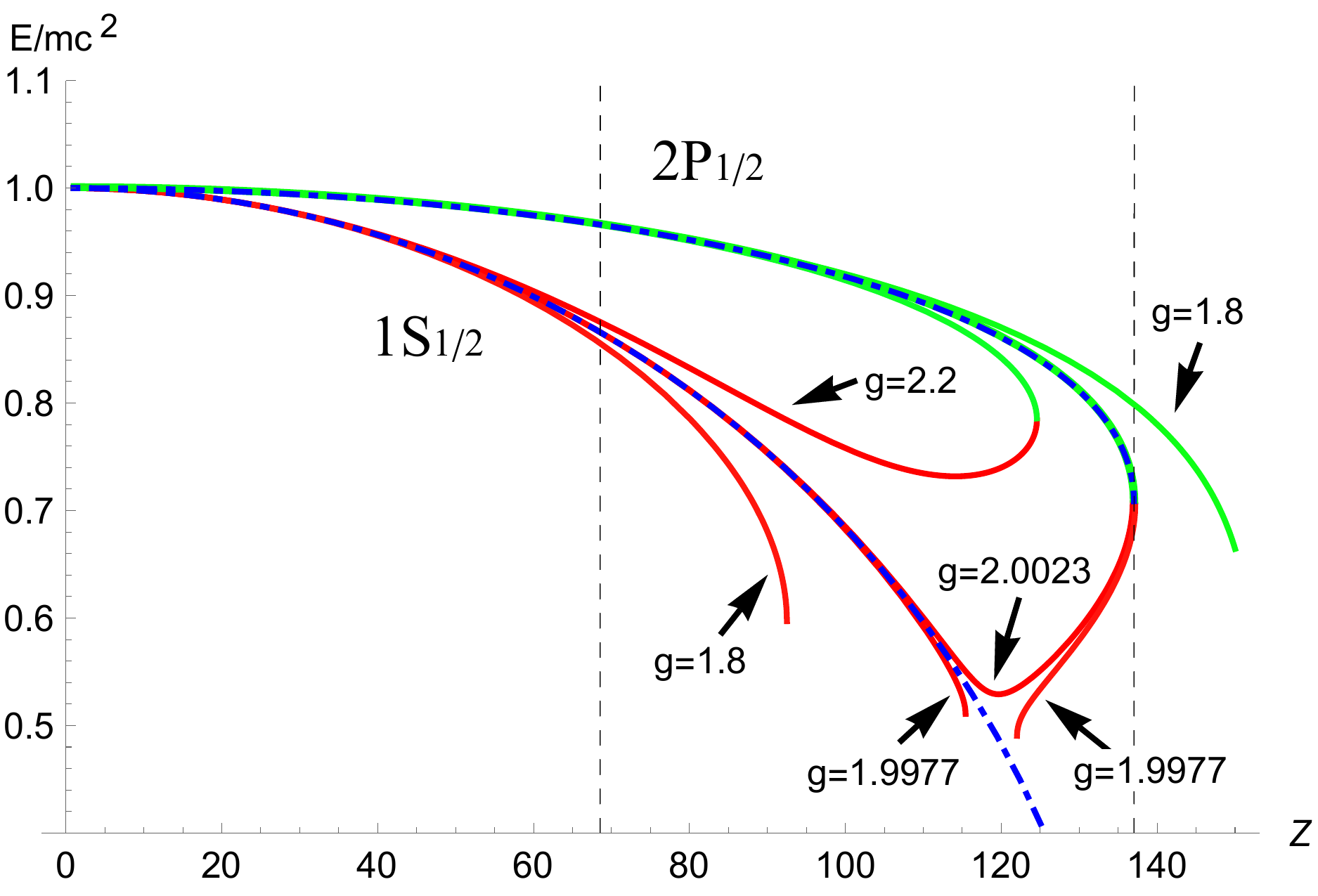}
 \caption[]{The KGP $1S_{1/2}$ (lower red curves) and $2P_{1/2}$ (upper green curves) energy levels for $g$-factor values $g~=~\{1.8,\ 1.9977,\ 2.0023,\ 2.2\}$ are shown for large $Z$ hydrogen-like atoms. The curves for the Dirac $g=2$ case for (lower dashed blue) $1S_{1/2}$ and (upper dashed blue) $2P_{1/2}$ are also presented.}
 \label{f03}
\end{figure}

This behavior can be seen in figure~\ref{f03}, which shows the meeting of the $1S_{1/2}$ and $2P_{1/2}$ states. For $|g|<2$ there is no state merging, but for small anomalies the solution is discontinuous in the sense that even for 1S$_1/2$ we see in figure~\ref{f03} a maximum allowed value of $Z$ at a finite energy. This behavior is reminiscent of the behavior we are familiar with for $1/r$ potential for the 2P$_1/2$ (seen in figure~\ref{f03}) and many other $g=2$ eigenstates. We know from study of numerical solutions of the Dirac equation that the regularization of the Coulomb potential by a finite nuclear size removes this singular behavior. It remains to be seen how this exactly works in the context of the KGP equation allowing for the magnetic anomaly.

The Dirac $g=2$ case acts as unique \lq\lq cusp\rq\rq\ point because even for very small anomalies, as emphasized in figure~\ref{f04}, the behavior of the states is strongly modified for the situation of high intensity Coulomb fields which are present in large $Z$ hydrogen-like nuclei.

Thaller~\cite{Thaller:1992ji} presented numerically computed DP equation energy levels for large $Z$ hydrogen like atoms. These numerical solutions involve crossings in energy levels between states with the same total angular quantum number $j$, but differing spin orientations such as $1S_{1/2}$ and $2P_{1/2}$; these states also have the behavior of diving into the antiparticle lower continuum even for $1/r$-potential. These features are not present for the KGP-Coulomb solution. However, there is a similarity between the numerical solutions of the DP equation and our analytical KGP solutions, because for $|g|>2$ the merging states as described above correspond to the crossing states in the DP solution.

The DP equation also allows for the so-called super-positronium states as described by Barut and Kraus~\cite{Barut:1975hz,Barut:1976hs}. Such states represent resonances due to the magnetic interaction that reside incredibly close to the center of the atom i.e $\sqrt{\langle r^{2}\rangle}\approx a\alpha\hbar/mc$, but this feature is absent from the KGP formation of the Coulomb problem as all KGP-Coulomb wave functions which can be normalized can be successfully matched to their Dirac ($g=2$) companions.

Because analytical solutions of DP equation, unlike our results for KGP, are not available it is hard to pinpoint precisely the origin of the diverse unpalatable behavior. However, we can hypothesize that the problems arise due to the pathological structure of DP equation where the magnetic anomaly rather than full magnetic moment appears. In any case we see that the Thaller solutions present pathologies quite akin to those we already described in our study of Landau energies. On the other hand KGP framework for large $Z$ shows some exciting and nice analytical behavior.

\section{Conclusions} \label{concl}
As we discussed in this work, there have been two natural extensions of the relativistic quantum mechanics to include anomalous magnetic moments; we also proposed a third one in section~\ref{IKGP}. Our study shows that description of magnetic moment in the context of relativistic quantum physics is today an unfinished subject. In view of the results presented we recognize that inherent to DP is a division of magnetic moment, a conserved quantity, into two different mathematical components. Readers who think that anything, but DP, is an invalid framework have to be at peace with this while rejecting a more natural unified description within the KGP dynamics.

Looking back at our results we highlight the comparison of magnetic moment dynamics of DP and KGP formulation of relativistic quantum mechanics. The DP equation breaks up the magnetic moment into an underlying spinor structure part inherent to the Dirac equation, and a dedicated anomalous part. In contrast, for the KGP, the entire effect of magnetic moment is contained in a single Pauli term irrespective of the magnetic moment\rq s size. We find that the two models disagree in their predicted energy levels for the homogeneous magnetic field and the Coulomb field.
\begin{itemize}
\item
For Landau levels both Eqs.\;\eqref{lan24b} and \eqref{lan25b} have the correct non-relativistic reduction at lowest order to Eq.\,\eqref{lan02}, the latter is much more unwieldy and obscures the physical interpretation. Moreover, in the relativistic limit the DP-Landau level energies Eqs.\,\eqref{lan02} and \eqref{lan25b} is that they include cross terms between the $g=2$ magnetic moment and anomalous terms in $a=(g-2)/2$, thus the result does not depend on the particle magnetic moment alone, there is a functional dependence on the magnetic anomaly $a$ as well. This is in our opinion not acceptable. On the contrary, for the KGP-Landau levels, Eq.\,\eqref{lan24b}, we have a simple dependence on the full magnetic moment $g$-factor. This simplicity allows, for the KGP equation, the straightforward analysis of physical systems and elegant expressions for their solutions. Thus Dirac\rq s beauty principle favors heavily the KGP considering the Landau levels.
\item
In the case of the Coulomb field there are even for weak fields measurable differences in both the Lamb shift and fine structure; the contribution to the Lamb shift and fine structure splitting are proportional to: KGP $g^{2}/8-1/2$, Eq.\,\eqref{lamb01} and $g^{2}/8$, Eq.\,\eqref{fs00} respectively, rather than: DP $g/2-1$ Eq.\,\eqref{lamb03}) and $g/2-1/2$, Eq.\,\eqref{fs02}) respectively.
\item For strong fields, both DP and KGP share the behavior of a shrinking particle/antiparticle gap for the ground state when $|g|>2$, though the expressions differ from each other, see section~\ref{sbl}). The models are extremely different in both their predicted energy values and the ultimate fate of the states as $B$ increases. For the KGP equation, the gap vanishes for in very strong magnetic fields Eq.\,\eqref{Bcrit}, and for DP the fields have to be even stronger. In both cases this behavior is unphysical and this issue can be resolved by the modification introduced as the IKGP equation Eq.\,\eqref{IKGP01} for homogeneous magnetic fields in section~\ref{IKGP}. The IKGP equation produces new phenomena which we hope to address separately in near future.
\item
For strong fields with large $Z$ hydrogen-like atoms, in the KGP model, there is a difficult to interpret merging of states that share the same total angular momentum, but differ in orbital angular momentum; the states lose physical meaning at the merging point (see figure~\ref{f03}). These are the same states that are problematic in the DP equation in that they cross, rather than merge. Either case indicates interesting physics involving these states.
\end{itemize}

This shows that both formulations of magnetic moment, DP and KGP, cannot be simultaneously correct and it is likely that one of these descriptions more closely describes nature. The quantum level splitting differences, which are on the order of tens of kHz for the hydrogen atom, are either already within reach or will soon be in reach of modern spectroscopic experiments. These differences, however, are not sufficient to explain mysteries such as the proton radius puzzle of Pohl~{\em et. al.}~\cite{Pohl:2013yb}. However, we did not study the relativistic scattering in KGP context which provides the other \lq end\rq\ of the puzzle. We also see further opportunity since there can be still better effective approaches characterizing point magnetic moment dynamics in the realm of RQM (section~\ref{IKGP}).

Because for the time being the magnetic moment interaction cannot be uniquely defined in quantum mechanics as discussed here, a behavior we also found in the classical dynamics, ~\cite{Rafelski:2017hce}, there is a strong motivation to continue the study of different models of this phenomenon until a convincing formulation emerges, or the experiment can settle which of the models better reflects reality. While the benefit of such study to atomic and nuclear systems is immediately obvious, there are ramification relevant to stellar astrophysics as discussed in sections~\ref{sbl} and \ref{IKGP} and cosmology through the study of (Dirac) neutrinos which are predicted \cite{Fujikawa:1980yx} to have a small, but yet undetected, magnetic moment. 

\subsubsection*{Acknowledgment} One of us (JR) thanks his colleagues Tobias Fischer and Ludwik Turko at the University of Wroclaw for fruitful discussions, and acknowledges support of his visits in Wroclaw by the Polish National Science Center under contract No. UMO-2016/23/B/ST2/00720\,.



\begin{thebibliography}{99}

\bibitem{Rafelski:2017hce} 
 J.~Rafelski, M.~Formanek and A.~Steinmetz,
 \lq\lq Relativistic dynamics of point magnetic moment,\rq\rq\ 
\href{https://doi.org/10.1140/epjc/s10052-017-5493-2}{\emph{Eur.\ Phys.\ J.\ C} {\bf 78}, 6 (2018)}
 \href{http://arxiv.org/abs/arXiv:1712.01825}{[arXiv:1712.01825 [physics.class-ph]]}.
 
\bibitem{Brown:1958zz} 
 L. M. Brown, 
 \lq\lq Two-component fermion theory.\rq\rq\ 
 \href{https://doi.org/10.1103/PhysRev.111.957}{\emph{Phys.\ Rev.} {\bf 111}, 957 (1958)}
 
\bibitem{Veltman:1997am} 
 M.~J.~G.~Veltman,
 \lq\lq Two component theory and electron magnetic moment,\rq\rq\ 
 \href{http://www.actaphys.uj.edu.pl/findarticle?series=Reg&vol=29&page=783}
 {Acta Phys.\ Polon.\ B {\bf 29}, 783 (1998)}
 \href{https://arxiv.org/abs/hep-th/9712216}{[arXiv:hep-th/9712216]}


\bibitem{Kaspi:2017fwg}
 V.~M.~Kaspi, and A.~Beloborodov, 
 \lq\lq Magnetars.\rq\rq\ 
 \href{https://doi.org/10.1146/annurev-astro-081915-023329}{\emph{Ann.\ Rev.\ Astron.\ Astrophys.} {\bf 55} (2017) 261 } \href{https://arxiv.org/abs/1703.00068}{[arXiv:1703.00068 [astro-ph.HE]]}

\bibitem{Rafelski:1976ts} 
 J. Rafelski, L. P. Fulcher, and A. Klein. 
 \lq\lq Fermions and bosons interacting with arbitrarily strong external fields.\rq\rq\ 
 \href{https://doi.org/10.1016/0370-1573(78)90116-3}{\emph{Phys.\ Rept.} {\bf 38}, 227 (1978)}


\bibitem{Greiner:1985ce} 
 W. Greiner, B. M\"uller, and J. Rafelski. 
 \href{https://doi.org/10.1007/978-3-642-82272-8}{\emph{Quantum Electrodynamics of Strong Fields}}
 (Springer-Verlag Berlin Heidelberg, 1985) 594 P. (Texts and Monographs In Physics) .


\bibitem{Rafelski:2016ixr}
 J.~Rafelski, J.~Kirsch, B.~M\"uller, J.~Reinhardt, and W.~Greiner,
 \lq\lq Probing QED vacuum with heavy ions.\rq\rq\ 
 \emph{New horizons in fundamental physics.} 
 \href{https://doi.org/10.1007/978-3-319-44165-8\_17}{FIAS Interdisc.\ Sci.\ Ser.\ 211-251 (Springer 2017)} \href{http://arxiv.org/abs/1604.08690}{[arXiv:1604.08690 [nucl-th]]}

\bibitem{Evans:2018kor}
 S.~Evans, and J.~Rafelski,
 \href{https://doi.org/10.1103/PhysRevD.98.016006}{ \emph{Phys.\ Rev.\ D} {\bf 98} (2018) no.1, 016006} \href{https://arxiv.org/abs/1805.03622}{[arXiv:1805.03622 [hep-ph]]}


\bibitem{Dunne:2014qda}
 G.~V.~Dunne,
 \lq\lq Extreme quantum field theory and particle physics with IZEST, \rq\rq\
\href{https://doi.org/10.1140/epjst/e2014-02156-4}{\emph{Eur.\ Phys.\ J.\ ST} {\bf 223} (2014) no.6, 1055.}

\bibitem{Hegelich:2014tda}
 B.~M.~Hegelich, G.~Mourou, and J.~Rafelski,
 \lq\lq Probing the quantum vacuum with ultra intense laser pulses, \rq\rq\ 
\href{https://doi.org/10.1140/epjst/e2014-02160-8}{\emph{Eur.\ Phys.\ J.\ ST} {\bf 223} (2014) no.6, 1093} \href{http://arxiv.org/abs/1412.8234}{ [arXiv:1412.8234 [physics.optics]]} 


\bibitem{Pohl:2013yb} 
 R.~Pohl, R.~Gilman, G.~A.~Miller, and K.~Pachucki,
 \lq\lq Muonic hydrogen and the proton radius puzzle,\rq\rq\ 
 \href{https://doi.org/10.1146/annurev-nucl-102212-170627}{\emph{Ann.\ Rev.\ Nucl.\ Part.\ Sci.} {\bf 63}, 175 (2013)} \href{https://arxiv.org/abs/1301.0905}{[arXiv:1301.0905 [physics.atom-ph]]}
 
 
\bibitem{Zemach:1956zz}
 A.~C.~Zemach,
 \lq\lq Proton Structure and the Hyperfine Shift in Hydrogen,\rq\rq
 \href{https://doi.org/10.1103/PhysRev.104.1771}{Phys.\ Rev.\ {\bf 104} (1956) 1771.}

\bibitem{Schwinger:1951nm}
 J.~S.~Schwinger,
\lq\lq On gauge invariance and vacuum polarization,\rq\rq
 \href{https://doi.org/10.1103/PhysRev.82.664}{ Phys.\ Rev.\ {\bf 82} (1951) 664.}


\bibitem{Tsai:1972iq}
 W.~Y.~Tsai, and A.~Yildiz,
\lq\lq Motion of charged particles in a homogeneous magnetic field.\rq\rq\ 
\href{https://doi.org/10.1103/PhysRevD.4.3643}{\emph{Phys.\ Rev.\ D} {\bf 4} (1971) 3643} 


\bibitem{Thaller:1992ji}
 B.~ Thaller. \href{https://doi.org/10.1007/978-3-662-02753-0}{\emph{The Dirac Equation}} 
 (Springer 1992) 357 p. (Texts and monographs in physics)

\bibitem{Barut:1975hz}
 A.~O.~Barut and J.~Kraus,
 \lq\lq Resonances in e+ e- system due to anomalous magnetic moment interactions.\rq\rq\ 
 \href{https://doi.org/10.1016/0370-2693(75)90696-6}{\emph{Phys.\ Lett.} {\bf 59B} (1975) 175 } 

\bibitem{Barut:1976hs} 
A.~O.~Barut and J.~Kraus, 
 \lq\lq Solution of the Dirac equation with Coulomb and magnetic moment interactions.\rq\rq\ \href{https://doi.org/10.1063/1.522932}{\emph{J.\ Math.\ Phys.} {\bf 17} (1976) 506} 


\bibitem{Knecht:2003kc}
 M.~Knecht,
 \lq\lq The anomalous magnetic moment of the muon: a theoretical introduction.\rq\rq\ 
 \emph{Lectures on Flavor Physics.} 
 \href{https://doi.org/10.1007/978-3-540-44457-2\_2}{\emph{Lect.\ Notes Phys.} {\bf 629} 37-84 (Springer 2004)} \href{https://arxiv.org/abs/hep-ph/0307239}{arXiv:hep-ph/0307239}

\bibitem{Fock:1937dy}
 V. Fock, 
 \lq\lq Proper time in classical and quantum mechanics.\rq\rq\ (In German) 
 Phys.\ Z.\ Sowjetunion {\bf 12} (1937) 404. 

\bibitem{Feynman:1958ty}
 R. P. Feynman and M. Gell-Mann,
 \lq\lq Theory of the Fermi Interaction\rq\rq\ 
 \href{https://doi.org/10.1103/PhysRev.109.193}{\emph{Phys.\ Rev.} {\bf 109} (1958) 193}
 
\bibitem{Cortes:1992wr}
 J.~L.~Cortes, J.~Gamboa and L.~Velazquez,
 \lq\lq Second order formalism for fermions.\rq\rq\ 
 \href{https://doi.org/10.1016/0370-2693(93)91198-V}{\emph{Phys.\ Lett.} B {\bf 313} (1993) 108} 
 \href{https://arxiv.org/abs/hep-th/9301071}{[arXiv:hep-th/9301071]}

\bibitem{AngelesMartinez:2011nt}
 R.~Angeles-Martinez and M.~Napsuciale,
 \lq\lq Renormalization of the QED of second-order spin 1/2 fermions.\rq\rq\ 
 \href{https://doi.org/10.1103/PhysRevD.85.076004}{\emph{Phys.\ Rev.\ D} {\bf 85} (2012) 076004} 
 \href{http://arxiv.org/abs/1112.1134}{[arXiv:1112.1134 [hep-ph]]}

\bibitem{DelgadoAcosta:2010nx}
 E.~G.~Delgado-Acosta, M.~Napsuciale and S.~Rodriguez,
 \lq\lq Second order formalism for spin 1/2 fermions and Compton scattering\rq\rq\ 
 \href{https://doi.org/10.1103/PhysRevD.83.073001}{\emph{Phys.\ Rev.} D {\bf 83} (2011) 073001}
 \href{http://arxiv.org/abs/1012.4130}{[arXiv:1012.4130 [hep-ph]]}

\bibitem{Feshbach:1958wv}
 H.~Feshbach and F.~Villars,
 \lq\lq Elementary relativistic wave mechanics of spin 0 and spin 1/2 particles,\rq\rq
 \href{https://doi.org/10.1103/RevModPhys.30.24}{\emph{Rev.\ Mod.\ Phys.\ } {\bf 30} (1958) 24.}

\bibitem{Robson:1996et}
 B.~A.~Robson, D.~S.~Staudte
 \lq\lq An eight-component relativistic wave equation for spin- particles I,\rq\rq
 \href{https://doi.org/10.1088/0305-4470/29/1/017}{\emph{J. Phys. A: Math. Gen.} {\bf 29}, 157 (1996)}

\bibitem{Staudte:1996ey}
 D.~S.~Staudte
 \lq\lq An eight-component relativistic wave equation for spin- particles II,\rq\rq
 \href{https://doi.org/10.1088/0305-4470/29/1/018}{\emph{J. Phys. A: Math. Gen.} {\bf 29}, 169 (1996)}


\bibitem{Johnson:1950zz}
 M.~H.~Johnson and B.~A.~Lippmann,
 \lq\lq Motion in a constant magnetic field.\rq\rq\ 
 \href{https://doi.org/10.1103/PhysRev.76.828}{\emph{ Phys.\ Rev.} {\bf 77} (1950) 702}. 

\bibitem{Ferrer:2009nq}
 E.~J.~Ferrer and V.~de la Incera,
 \lq\lq Dynamically generated anomalous magnetic moment in massless QED,\rq\rq\ 
 \href{https://doi.org/10.1016/j.nuclphysb.2009.08.024}{\emph{Nucl.\ Phys.\ B} {\bf 824} (2010) 217}.
 \href{http://arxiv.org/abs/0905.1733}{[arXiv:0905.1733 [hep-ph]]}

\bibitem{Rafelski:2012ui}
 J.~Rafelski and L.~Labun,
 \lq\lq A Cusp in QED at g=2,\rq\rq\ arXiv print (2012) 
\href{https://arxiv.org/abs/1205.1835}{arXiv:1205.1835 [hep-ph]}. 

\bibitem{r61}
 M. E. Rose. 
 \emph{Relativistic electron theory.}
 (John Wiley \& Sons 1961). p.51.

\bibitem{b69}
 G. A. Baym.
 \emph{Lectures on Quantum Mechanics.}
 (Westview Press, 1969). p.526. 

\bibitem{Martin:1958zz}
 P.~C.~Martin and R.~J.~Glauber,
 \lq\lq Relativistic Theory of Radiative Orbital Electron Capture,\rq\rq\ 
 \href{https://doi.org/10.1103/PhysRev.109.1307}{\emph{Phys.\ Rev.} {\bf 109} (1958) 1307}
 
\bibitem{bi62}
 L. C. Biedenharn. 
 \lq\lq Remarks on the relativistic Kepler problem,\rq\rq\ 
 \href{https://doi.org/10.1103/PhysRev.126.845}{\emph{Phys. Rev.} {\bf 126} (1962) 845}

\bibitem{Niederle:2004bx}
 J.~Niederle and A.~G.~Nikitin,
 \lq\lq The relativistic Coulomb problem for particles with arbitrary half-integer spin,\rq\rq\ 
 \href{https://doi.org/10.1088/0305-4470/39/34/023}{\emph{J.\ Phys.\ A} {\bf 39} (2006) 10931}.
 \href{https://arxiv.org/abs/hep-th/0412214}{[arXiv:hep-th/0412214]}

\bibitem{iz80}
 C. Itzykson, and J-B. Zuber. 
 \emph{Quantum field theory.} 
 (McGraw-Hill, 1980). p.360.
 

\bibitem{sy90}
 J. R. Sapirstein, and D. R. Yennie, 1990, in \emph{Quantum Electrodynamics},
 edited by T. Kinoshita. World Scientific, chapter 12, pp. 560–672.

\bibitem{Jentschura:1996zz}
 U. Jentschura, and K. Pachucki 
 \lq\lq Higher-order binding corrections to the Lamb shift of 2P states,\rq\rq\ 
\href{https://doi.org/10.1103/PhysRevA.54.1853}{\emph{Phys.\ Rev.\ A} {\bf 54} (1996) 1853}
 
\bibitem{Eides:2000xc}
 M.~I.~Eides, H.~Grotch and V.~A.~Shelyuto, 
 \lq\lq Theory of light hydrogenlike atoms,\rq\rq\ 
 \href{https://doi.org/10.1016/S0370-1573(00)00077-6}{\emph{Phys.\ Rept.} {\bf 342} (2001) 63}.
 \href{https://arxiv.org/abs/hep-ph/0002158}{[arXiv:hep-ph/0002158]}


\bibitem{Mohr:2012tt} 
 P.~J.~Mohr, B.~N.~Taylor and D.~B.~Newell, 
 \lq\lq CODATA recommended values of the fundamental physical constants: 2010,\rq\rq\ 
 \href{https://doi.org/10.1103/RevModPhys.84.1527}{\emph{Rev.\ Mod.\ Phys.} {\bf 84} (2012) 1527}
 \href{https://arxiv.org/abs/1203.5425}{[arXiv:1203.5425 [physics.atom-ph]]}

\bibitem{Jancovici:1970ep} 
 B.~Jancovici,
 \lq\lq Radiative correction to the ground-state energy of an electron in an intense magnetic field,\rq\rq
 \href{https://doi.org/10.1103/PhysRev.187.2275}{\emph{Phys.\ Rev.\ } {\bf 187}, 2275 (1969).}

\bibitem{Ferrer:2015wca}
 E.~J.~Ferrer, V.~de la Incera, D.~Manreza Paret, A.~Pérez Martínez and A.~Sanchez,
 \lq\lq Insignificance of the anomalous magnetic moment of charged fermions for the equation of state of a magnetized and dense medium,\rq\rq
 \href{https://doi.org/10.1103/PhysRevD.91.085041}{\emph{Phys.\ Rev.\ D} {\bf 91} (2015) no.8, 085041}
 \href{https://arxiv.org/abs/1501.06616}{[arXiv:1501.06616 [hep-ph]]}.

\bibitem{Broderick:2000pe}
 A.~Broderick, M.~Prakash and J.~M.~Lattimer,
 \lq\lq The Equation of state of neutron star matter in strong magnetic fields,\rq\rq
 \href{https://doi.org/10.1086/309010}{\emph{Astrophys.\ J.\ } {\bf 537} (2000) 351}
 \href{https://arxiv.org/abs/astro-ph/0001537}{[astro-ph/0001537]}.


\bibitem{Labun:2008re}
 L.~Labun and J.~Rafelski,
\lq\lq Vacuum Decay Time in Strong External Fields,\rq\rq\
\href{https://doi.org/10.1103/PhysRevD.79.057901}{\emph{Phys.\ Rev.\ D} {\bf 79} (2009) 057901}
\href{https://arxiv.org/abs/0808.0874}{[arXiv:0808.0874 [hep-ph]]} 

\bibitem{Fujikawa:1980yx} 
 K.~Fujikawa and R.~Shrock,
 \lq\lq The Magnetic Moment of a Massive Neutrino and Neutrino Spin Rotation,\rq\rq\
 \href{https://doi.org/10.1103/PhysRevLett.45.963}{\emph Phys.\ Rev.\ Lett.\ {\bf 45} (1980) 963.}

\end{thebibliography}
\end{document}